\documentclass[%
 aip,
 jmp,%
 amsmath,amssymb,
 reprint,%
]{revtex4-1}

\usepackage{graphicx}
\usepackage{dcolumn}
\usepackage{bm}

\begin{document}

\preprint{AIP/123-QED}

\title[Lattice models for protein organization throughout thylakoid membrane stacks ]{Lattice models for protein organization throughout thylakoid membrane stacks}

\author{Andreana M. Rosnik} 
\author{Phillip L. Geissler}
\email{geissler@berkeley.edu}
\affiliation{Department of Chemistry, University of California, Berkeley, California 94720}
\affiliation{Molecular Biophysics and Integrated Bioimaging Division, Lawrence Berkeley National Lab, Berkeley, California 94720}

\maketitle 

\section{Abstract}
Proteins in photosynthetic membranes can organize into patterned 
arrays that span the membrane's lateral size. 
Attractions between proteins in different layers of a membrane stack 
can play a key role in this ordering, as was suggested by microscopy 
and fluorescence spectroscopy 
and demonstrated by computer simulations of a coarse-grained model.
  The architecture of thylakoid membranes, however, also provides opportunities 
  for inter-layer interactions that
  instead disfavor the high protein densities of ordered arrangements.
  Here we explore the interplay between these opposing driving forces,
  and in particular the phase transitions that emerge in the periodic
  geometry of stacked thylakoid membrane discs.
  We propose a lattice model that roughly accounts for proteins'
  attraction within a layer and across the stromal gap, steric
  repulsion across the lumenal gap, and regulation of protein density
  by exchange with the stroma lamellae.
  Mean field analysis and computer simulation reveal rich phase
  behavior for this simple model, featuring a broken-symmetry striped
  phase that is disrupted at both high and low extremes of 
  chemical potential.
  The resulting sensitivity of microscopic protein arrangement
  to the thylakoid's mesoscale vertical structure raises intriguing
  possibilities for regulation of photosynthetic function.

\section{Statement of Significance}
This work develops the first theoretical model for grana-spanning
spatial organization of photosynthetic membrane proteins. Based on the
stacked-disc structure of thylakoids in chloroplasts, it focuses on a
competition between interactions that dominate in different parts of
the granum. Analysis and computer simulations of the model reveal
striped patterns of high protein density as a basic consequence of
this competition, patterns that acquire long-range order for a broad
range of physical conditions. Because natural changes in light and
stress conditions can substantially alter the strengths of these
competing interactions, we expect that an ordered phase with
periodically modulated protein density is thermodynamically stable at
or near some physiological conditions.

\section{Introduction} \label{intro}

Photosynthetic membranes are dense in proteins that cooperate to
execute the complicated chemistry fundamental to light-harvesting and
other components of photosynthesis. Membrane functionality depends not only on
these proteins, but also supramolecular
spatial arrangements thereof. Both the architecture of the membranes and the
interactions of the protein components influence protein organization. Both levels of complexity are further influenced by
light and physiological conditions.

\begin{figure}[h]
\includegraphics[width=0.75\textwidth]{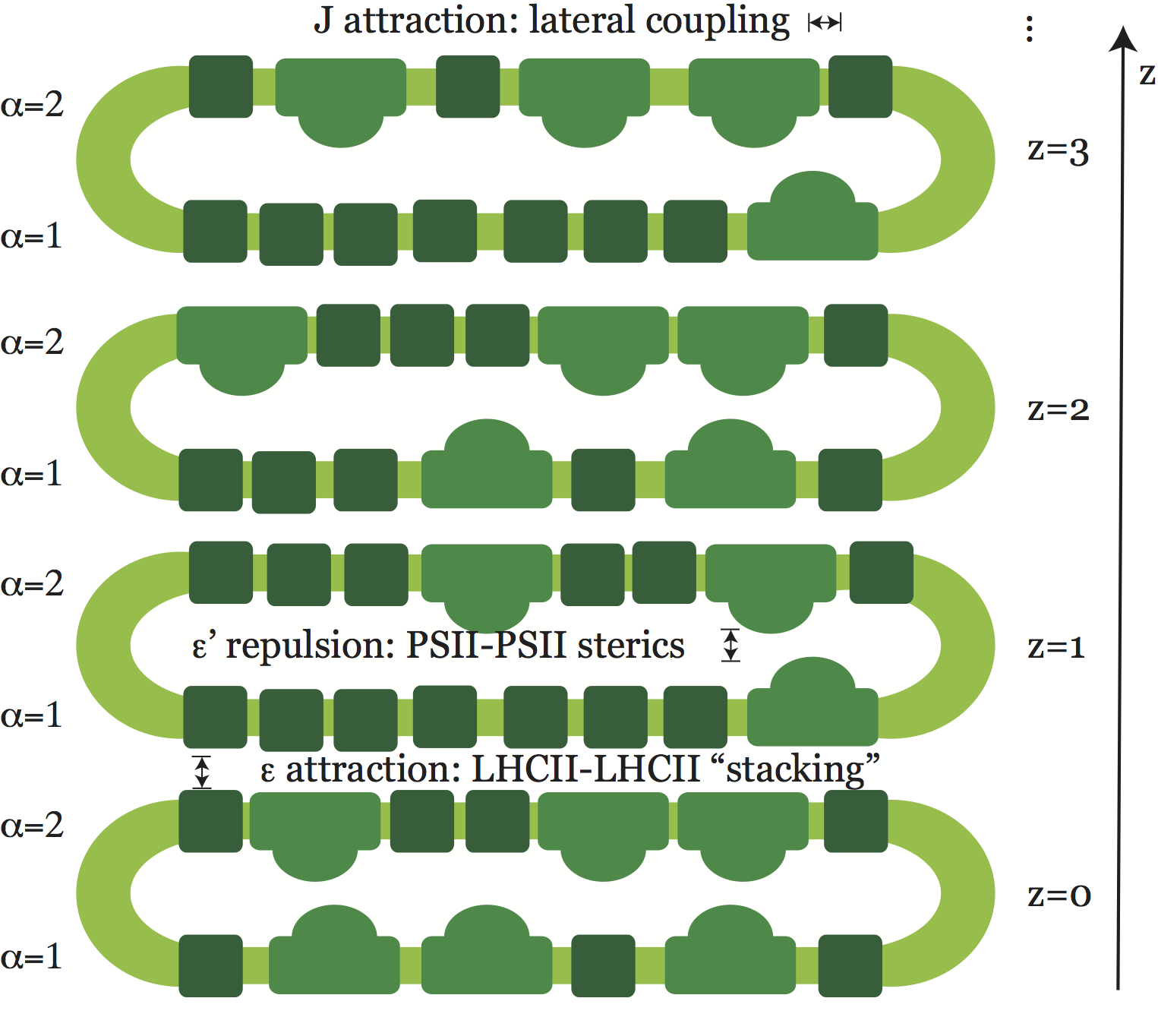}
\caption{\label{fig:explain} Schematic cross-section of a short stack
  of thylakoids discs. Dark green squares represent LHCII molecules,
  lighter green domed shapes represent PSII, and yellow-green bands
  represent lipid bilayers. 
  Each disc (indexed by an integer $z$) comprises two layers (indexed
  $\alpha=1$ and $\alpha=2$). Protein attraction within each layer is
  assigned an energy scale $J$ in our lattice model. Aligned LHCIIs
  in subsequent layers can engage in favorable stacking interactions,
  which is assigned an energy $\epsilon$ in the model. Protrusion of
  PSII into the lumen spaces (\textit{i.e.}, the interior of each disc) may
  lead to steric repulsion between the two layers of each
  disc. Mediated by thylakoid gap and membrane fluctuations, the
  effective steric energy scale is denoted $\epsilon'$.}
\end{figure}

In higher plants, photosynthetic membranes are
arranged as stacks (called grana) of discs (called thylakoids). Each
thylakoid, measuring roughly 300-600 nm in diameter and 10-15 nm
thick, is bounded above and below by a lipid bilayer densely populated
with photosynthetic proteins (See Fig.~\ref{fig:explain}). \cite{DekkerBoekema2005ar,PribilLabsLeister2014ar,NevoCharuviTsabariReich2012ar}
A typical granum is composed of 10-100 thylakoid discs, spaced
vertically by 2-4 nm.
Grana do not exist in isolation in chloroplasts; rather, they are
connected by unstacked membranes called stroma lamellae, which tend to
be longer and have different protein composition. \cite{DekkerBoekema2005ar,NevoCharuviTsabariReich2012ar,Daum2011} 
See Refs. \cite{DekkerBoekema2005ar,NevoCharuviTsabariReich2012ar} 
for visual representations of the membrane architecture.

This intricate geometry provides diverse opportunities for 
association among transmembrane proteins. We focus exclusively on 
protein-protein interactions. While lipids are an important component of the 
protein environment, our models do not resolve their structure or 
composition explicitly. The role of lipids is limited in our perspective to 
an implicit influence on the strength of protein-protein attraction. The 
character of protein-protein interactions we have in mind -- sharply 
repulsive at short range and smoothly attractive out to distances of ~1 nm -- 
is not special. Indeed, most folded proteins engage in similar forces, whether 
free in solution or bound within a membrane. The consequence of these
 interactions, however, may be strongly influenced by the geometry 
 of the thylakoid. We emphasize in particular molecular organization 
 involving two particular proteins, photosystem II (PSII) and 
 light-harvesting complex II (LHCII), which abound in the central, 
 mostly flat portion of thylakoid discs. \cite{DekkerBoekema2005ar,QinSugaKuangShen2015ar,WeiSuCao2016ar,LiguoriPerioleMarrinkCroce2015ar}
``Super-complexes" comprising a handful of these proteins can form
with a variety of ratios
LHCII:PSII. \cite{DekkerBoekema2005ar,NosekKouril2017ar}
Super-complexes are situated within a single lipid bilayer, but 
their stability may be influenced by interactions across 
the gap separating distinct thylakoid discs.
 \cite{SchneiderGeissler2013ar,ChowKimHortonAnderson2005ar,KirchhoffRogner2007ar} 
These interactions across the gap appear to be net attractive due to 
solvent mediation of interactions between polar, protruding domains of LHCII proteins.

Such attractive ``stacking''
interactions may also drive larger scale organization of these
proteins 
within the plane of the bilayer, forming laterally
extended periodic arrays that have been
observed. \cite{ChowKimHortonAnderson2005ar,StandfussKuhlbrandt2005ar,PhuthongHuangGrossman2015ar,OnoaSchneider2014ar,RubanJohnson2015ar} Computational work has suggested that these
lateral arrays signal a phase transition to a crystalline state that would
exhibit truly long-range two-dimensional order in the absence of constraints on
protein population and disc size. \cite{SchneiderGeissler2013ar,SchneiderGeissler2014ar,LeePaoSmit2015ar,Amarnath2015ar}
Small changes in protein
composition, density, and interaction strength could thus trigger
sudden large-scale reorganization. Diminished stacking during state
transitions and non-photochemical quenching (NPQ) processes (processes of thylakoid restructuring to shift electronic excitations or to minimize photo-oxidative damage, respectively) 
may reflect control mechanisms that
exploit this sensitivity. \cite{EricksonWakaoNiyogi2015ar} 

In bulk solution, protein-protein attractions can stabilize a 
state of uniformly high density extending in all directions. 
By contrast, stacking interactions in the thylakoid, by themselves, 
do not correlate protein density across large vertical distances. 
Because LHCII lacks domains suitable to propagate such an 
interaction over the 4-6 nm inner thylakoid gap, stacking 
attractions do not act between the two membrane layers 
comprising the same disc. Only adjacent membrane 
layers of consecutive discs are influenced by this attraction, 
limiting the range of stacking-induced correlations to a 
microscopic scale in the vertical direction. Development 
of extended order in the thylakoid geometry would require additional factors.

Vertical protein-protein interactions in a stack of thylakoids 
can also be repulsive in character. Due to narrow spacing between 
apposed membranes, and the significant protrusion of certain 
proteins into the region between stacked membranes, steric repulsion 
is likely to influence spatial organization in some circumstances. 
PSII in particular extends large domains towards the interior of 
thylakoid discs (called the lumen), a space that contracts under 
low light conditions. With sufficient contraction of the lumen, 
PSII molecules inhabiting a disc’s opposing membranes 
would be unable to share the same lateral position.
 \cite{ChowKimHortonAnderson2005ar,KirchhoffHallWood2011ar,Kirchhoff2014ar,Albertsson1982ar} 
Whereas in bulk solution steric repulsions primarily constrain 
the local packing of proteins in states of high density, in the 
thylakoid they can prevent the two layers of a disc from 
being simultaneously occupied at high protein density. As 
in the case of stacking attractions, these PSII repulsions can 
by themselves correlate protein density only over a limited 
vertical scale in the thylakoid geometry. PSII does not extend 
large domains into the space between discs, so that steric 
constraints likely operate only within each disc and only in 
light conditions that favor a small lumenal gap. Development 
of extended order from steric repulsion along the thylakoid 
stack would require additional factors. The consequences of 
such a constraint on protein organization, \textit{e.g.}, its 
implications for the stability of stacked protein arrays, have 
not been directly explored in either experiment or simulation. 
However, the implications of these spatial constraints on 
the diffusion of molecules in the lumen has been 
addressed in Refs. \cite{KirchhoffHallWood2011ar,Kirchhoff2014ar}.

For a system in low-light to dark conditions, stacking attractions 
across the inter-disc gap are strong, and steric repulsions 
across the lumenal gap can significantly constrain protein organization. 
These two interactions conflict in the thylakoid geometry, 
preventing a conventional scenario of aggregation that 
would be expected for such proteins in bulk solution. A state 
of vertically extended correlation among thylakoid proteins 
would instead involve spatially modulated order. To our 
knowledge this possibility has not yet been carefully explored, 
due in part to the challenges posed by highly structured 
biological membranes for microscopy as well as for molecular simulation.

Our work addresses the interplay between attractive and 
repulsive protein-protein forces within grana stacks. To date only 
one study has attempted to quantify the competition between 
attractive and repulsive protein-protein forces within grana stacks, 
and its sensitivity to changing physiological conditions. \cite{Puthiyaveetil2017} 
Different interactions likely prevail in different parts of the stack, 
due to proteins’ well-defined orientation relative to the lumen. 
We therefore focus on the possibility of spatially modulated order, 
patterns of protein density that alternate along the direction of stacking. 
To date such patterns have not been observed in experiment, 
but potential impacts of related kinds of granum-scale order 
on photosynthetic function have recently been discussed. \cite{Capretti2019ar} 
The computational works mentioned above utilize coarse-grained 
particle approaches, whereas our model 
will be a lattice-based approach; in this manner, we attempt to create the simplest possible 
explanatory model for vertical ordering in photosynthetic membranes.  \cite{SchneiderGeissler2013ar,SchneiderGeissler2014ar,LeePaoSmit2015ar,Amarnath2015ar}
Lattice models have been used for myriad studies of lipid organization in bilayers, 
as well as for lipid-protein membrane systems. \cite{Speck2010ar,MachtaVeatch2012ar,NajiBrown2007ar,Schick2012ar,PutzelSchick2008ar,FrazierAlmeida2007,YethirajWeisshaar2007,SchickKeller2008,MachtaVeatch2011,MitraSethna2018,MeerschaertKelly2015,HoshinoKomuraAndelman2015ar,HoshinoKomuraAndelman2017ar}

There is empirical evidence for vertically extended order within a
stack of membranes, though in a much simpler context 
and with a focus on lipids rather than proteins.
Specifically,
synthetic membrane systems, devoid of proteins, have been constructed
to examine compositional ordering in an array of lipid bilayers
with multiple lipid constituents. \cite{Tayebi2012ar,Tayebi2013ar}
Spatial modulations in lipid composition were observed to align and
extend throughout the entire membrane stack, establishing a basic
plausibility for the ordered phases discussed in this paper.

In order to examine the basic physical requirements for protein
correlations spanning an entire stack of thylakoids, we develop
minimal models that account for locally fluctuating protein
populations in a granum-like geometry. As described in Sec.~\ref{model}, these
fluctuations are biased by protein-dependent attractions between 
discs, and by steric
repulsion between proteins that reside in the same disc.
The strengths of
these interactions are determined by parameters that roughly represent
light conditions and protein phosphorylation states.
Such minimally detailed lattice models allow a thorough
statistical mechanics analysis that would not be possible with a more
elaborate representation. They also highlight symmetries that are
shared by more explicit models but difficult to recognize
amid atomistic details. Spontaneous breaking of such symmetries lies
at the heart of phase transitions, it has clear-cut signatures in
computer simulations, and it has important implications for
thermodynamic properties and microscopic correlations as phase
boundaries are approached. Finally, by sparing details we can potentially
address a range of different physiological situations whose essential
ordering scenarios coincide.

Using methods of
Monte Carlo simulation detailed in Sec.~\ref{methodmc}, as well as mean field
theories presented in Sec.~\ref{methodmft}, we find that strongly cooperative
behavior
of our lattice models
emerges over a wide range of conditions. As parameter values
are changed, the model system can cross phase boundaries where
intrinsic symmetries are spontaneously broken or restored. The
correspondingly sudden changes in the microscopic arrangement of
photosynthetic proteins suggest a mechanism for switching sharply
between distinct states of light harvesting activity, as discussed in
Sec.~\ref{discuss}. In Sec.~\ref{conclusion} we conclude. 

\section{Model}\label{model}

\subsection{Physical description}\label{physmodel}

Our model of stacked thylakoid discs
elaborates the familiar lattice
gas model of liquid-vapor phase transitions. We represent the
microscopic arrangement of proteins on a cubic lattice, resolving
their transiently high number density in some parts of the membrane
and low density in others. Proteins' specific identities and internal
structures are not resolved here; in discretizing space at the scale
of a protein diameter, we have notionally averaged out such details.
Our fluctuating degrees of freedom are thus binary variables $n$ for
each lattice site, indicating the local scarcity ($n=0$) or abundance
($n=1$) of protein. We refer to the local states
$n=0$ and $n=1$ as
unoccupied and occupied, respectively, although they do not strictly
indicate the presence of an individual molecule.

  The net protein density in our model
membranes may be specified explicitly, or else set indirectly through
a chemical potential $\mu$ that regulates density fluctuations. For
mathematical convenience we take the latter approach in most
calculations, in effect allowing exchange of material with a
reservoir.  Here, the stroma lamellae -- unstacked regions of
photosynthetic membrane -- could play the role of reservoir. For a
real thylakoid stack, such exchange might be very slow, so that the
largest-scale features of experimentally observed states are not well
equilibrated. In that case, the phase transitions discussed below
would not be manifested by coherent long-range order; instead,
internally ordered domains would appear and grow, but their size would
not reach a disc's entire lateral area.

Interaction energies are assigned wherever adjacent sites on the
lattice are occupied. The sign and strength of such an interaction
depends on the locations of the two lattice sites involved, as
depicted in Fig.~\ref{fig:explain}. Within a planar layer of the stack (a disc
comprises two layers), neighboring occupied sites contribute an
attractive energy $-J$, representing lateral forces of protein-protein
association. Stacking interactions occur between laterally aligned
sites on the facing layers of sequential discs in the granum; each
pair of occupied stacked sites contributes an attractive energy
$-\epsilon$.

Laterally aligned sites within the same disc are subject to a
repulsive energy $\epsilon'$, representing steric forces between
transmembrane proteins protruding into the lumen. The harshly
repulsive nature of steric interactions suggests that $\epsilon'$
should be very large, effectively enforcing a constraint of volume
exclusion.
 For this reason, we will consider $\epsilon'=\infty$ as a
special case. Termed the \textit{hard constraint limit}, this case offers
mathematical simplification as well as transparent connections to a
related class of spin models. Smaller values of $\epsilon'$, however,
may be more appropriate in situations where steric overlap can be
avoided through modest deformation of the membrane layers. Under high
light conditions, when thylakoid discs swell in the vertical
direction, very slight membrane deformation (or perhaps none at all)
could be sufficient to allow simultaneous occupation of laterally
aligned sites, corresponding to very small $\epsilon'$.

\begin{figure}
\centering
\includegraphics[width=0.85\textwidth]{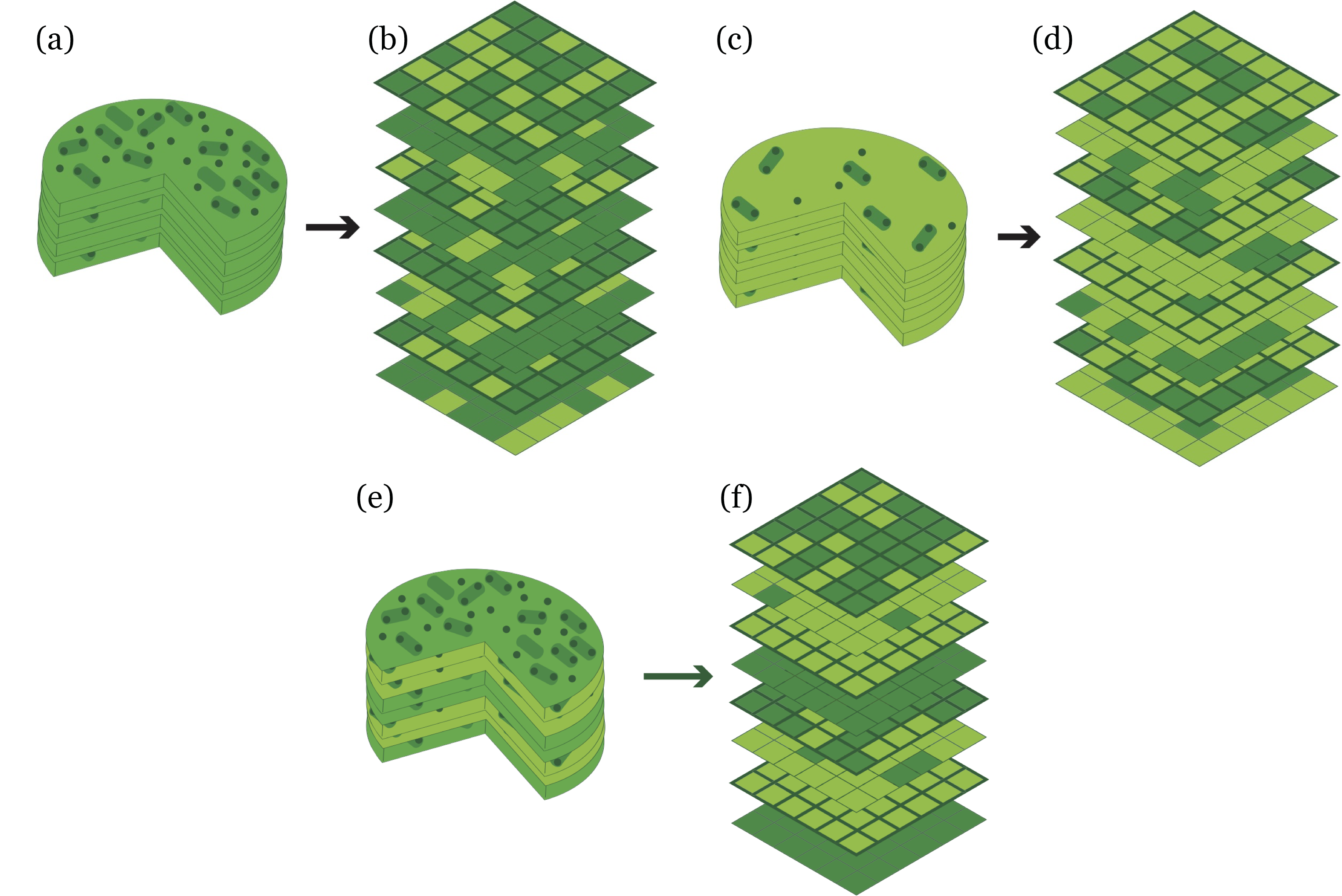}
\caption{\label{fig:maps}
Illustrations of three different granum states ((a), (c), and (e)), and
their representations in our lattice model ((b), (d), and (f)). In (a), (c), and
(e), yellow-green indicates membrane that is not inhabited by protein;
small dark circles are LHCII trimers; and oblong green shapes with
small circles are PSII-LHCII supercomplexes. In (b), (e), and (f),
yellow-green indicates a local sparsity of proteins, and dark green
represents a region that is densely populated by either protein. These
colors and shapes are used consistently throughout the paper. (a) and (b)
depict a state of high average protein density in the granum. (c) and (d)
depict a state of low average protein density. (e) and (f) illustrate a granum state with
 striped order. Here, layers of high and low protein density alternate
 vertically with a period of two discs. Each disc
 includes one high-density layer and one low-density layer; and each
 high-density layer is vertically adjacent to a dense layer on an
 adjacent disc. }
\end{figure}

The ground state of this model depends on values of the energetic
parameters $\mu$, $\epsilon$, $J$, and $\epsilon'$. Large, positive
$\mu$ encourages
the presence of proteins
and thus favors a high average value
$\bar{n}$ of the local occupation variable. In the limit
$\mu\rightarrow +\infty$, a state of complete occupation is thus
energetically minimum. At high
but finite
$\mu$ we generally expect thermodynamic
states that are densely populated with protein, 
as depicted in Fig.~\ref{fig:maps} (a) and (b). 
Conversely, at very negative values of $\mu$
we expect very sparse equilibrium states, as depicted in Fig.~\ref{fig:maps} (c) and (d).

Equilibrium states at modest $\mu$ are characterized by competition
among steric repulsion and the favorable energies of stacking and
in-plane association. Large $\epsilon'$ harshly penalizes lattice
states that are more than half full -- states which must feature
simultaneous occupation of laterally aligned sites within the same
disc. In order to realize in-plane attraction at half filling, one
layer of each thylakoid must be depleted of protein. The stack then
comprises a series of sparse and dense layers. Extensive stacking
interaction between discs requires a coherent sequence of these
layers, yielding ground states that are striped with a period of four
layers. This pattern is illustrated in Fig.~\ref{fig:maps} (e) and (f), and is quantified by an 
order parameter $\Delta n$ that compares protein density in the 
two layers of each thylakoid. More specifically, $\Delta n$ is a linear 
combination of layer densities, whose coefficients change sign 
with the same periodicity as the stripe pattern described. 

Macroscopically ordered stripes of protein density may be an unlikely
extreme in real grana.
Slow kinetics of protein exchange with stroma lamellae,
imperfect
grana architecture, or insufficiently strong interactions could all
prevent long-range coherence in practice. The tendency towards
ordering for dark to low light conditions can still be of importance,
\textit{e.g.}, in the form of transient striping over substantial length scales
or a steep decline in the population of vertically adjacent PSIIs
as the transition is approached.

The two layers of each disc are completely equivalent in our model
energy function. Stripe patterns, which populate the two layers
differently with a persistent periodicity, do not possess this
symmetry. Equilibrium states with $\Delta n \neq 0$ therefore require
a spontaneous symmetry breaking and a macroscopic correlation length,
and they must be separated from symmetric states by a phase boundary.
The computational and theoretical work reported in the following
sections aims to determine what, if any, thermodynamic conditions
allow for such symmetry-broken, coherently striped states at
equilibrium.

Our goals in exploring this model are both explanatory and functional.
For grana at conditions that have been examined in experiment,
we aim
to explain observed trends in protein distribution based on simple,
physically realistic
interactions.  Some interesting behaviors of our model, however, may
be fully realized only at protein densities that are challenging to
reach with existing experimental techniques. For these conditions, our
results offer predictions for the emergence of spatial patterns that
have not yet been observed. Possible physiological consequences of
this organization will be discussed in Sec.~\ref{discuss}.

\subsection{Mathematical definition}\label{mathmodel}
In order to describe quantitatively the energetics and ordering
described above, we introduce: a label $z$ distinguishing different
thylakoid discs, a label $\alpha$ distinguishing the two sides of
each disc, and a label $i$ distinguishing positions within each membrane
layer.
The notation $n_{\alpha,i}^{(z)}$ for an occupation variable therefore
indicates (a) the thylakoid disc to which it belongs, specified by a vertical
coordinate $z$ ranging from 1 to $L_z$, (b) which layer of the disc
it inhabits, $\alpha=1$ (bottom) or $\alpha=2$ (top), and (c) its lateral
position, specified by an integer $i$ ranging from 1 to $L_x L_y$.
(See Fig.~\ref{fig:explain}). Density and striping order parameters are
then defined as
\begin{equation} \label{eq:ordern}
\bar{n} \equiv (2 L_x L_y L_z)^{-1} \sum_{z,i,\alpha} n_{\alpha,i}^{(z)}
\end{equation}
and
\begin{equation} \label{eq:orderdn}
  \Delta n \equiv (2 L_x L_y L_z)^{-1} \sum_{z,i}
(-1)^z
  (n_{1,i}^{(z)} - n_{2,i}^{(z)}),
\end{equation}
and the total energy of a configuration $\{n_{\alpha,i}^{(z)} \}$ is
written
\begin{eqnarray} \label{eq:ham}
  H[\{n_{\alpha,i}^{(z)} \}] &&=  - \mu \sum_{z,\alpha} \sum_i n_{\alpha,i}^{(z)}
  - J \sum_{z,\alpha} \sum_{i,j}{}^{'} n^{(z)} _{\alpha,i} n^{(z)} _{\alpha,j}
  \nonumber\\
  &&- \epsilon \sum_{z} \sum_{i} n^{(z)} _{2,i} n^{(z+1)} _{1,i}
  + \epsilon^{'} \sum_{z} \sum_{i} n_{1,i}^{(z)} n_{2,i}^{(z)},
\end{eqnarray}
where the primed summation extends over distinct pairs of lateral
nearest neighbors. As described above, each occupation variable
$n_{\alpha,i}^{(z)}$ adopts values 1 (occupied) or 0 (unoccupied). The
energetic parameters $\epsilon$ (in-plane attraction), $J$ (stacking
attraction), and $\epsilon'$ (steric repulsion) are all positive
constants. At temperature $T$, the equilibrium probability
distribution of $\{n_{\alpha,i}^{(z)} \}$ is proportional to the
Boltzmann weight $e^{-\beta H}$, where $\beta \equiv 1/ k_B T$.

In addition to transparent spatial symmetries, this model possesses a
symmetry with respect to inverting occupation variables. Applying the
transformation $\hat{n}_{\alpha,i}^{(z)} = 1-n_{\alpha,i}^{(z)}$ to
all lattice sites generates from any configuration
$\{n_{\alpha,i}^{(z)} \}$ a dual configuration
$\{\hat{n}_{\alpha,i}^{(z)} \}$ whose probability is also generally
different from the original. As in the lattice gas, a certain choice
of parameters renders the Boltzmann weight invariant under this
transformation. In our case this statistical invariance occurs when $-
2\mu - 4 J - \epsilon + \epsilon'=0$, establishing a line of symmetry
in parameter space. More usefully for our purposes, the duality
establishes pairs of equilibrium states with related thermodynamic
properties. Specifically, the states $(\mu,\epsilon,J,\epsilon',T)$
and $(\hat{\mu},\epsilon,J,\epsilon',T)$ have identical statistics of
$\Delta n$ for the choice 
\begin{eqnarray} \label{eq:map}
\hat{\mu} = -\mu-4 J - \epsilon + \epsilon'
\end{eqnarray} 
Viewing density rather than chemical potential as a control parameter,
distributions of $\Delta n$ are identical in pairs of thermodynamic
states $(\bar{n},\epsilon,J,\epsilon',T)$
and $(\hat{\bar{n}},\epsilon,J,\epsilon',T)$ related by
$\hat{\bar{n}} = 1-\bar{n}$; in other words, $\bar{n}=1/2$ is also
a line of symmetry due to duality.

For the phase transitions of interest here, these arguments guarantee
that any phase boundary at chemical potential $\mu$ (or density
$\bar{n}$) is mirrored by a dual transition at $\hat{\mu}$ (or
$\hat{\bar{n}}$), for any consistent choice of $\epsilon,J,\epsilon'$,
and $T$. More physically, any phase change induced by controlling
protein density must exhibit reentrance (or else occur exactly at the
line of symmetry, which we do not observe).

In simpler terms, imagine an initial equilibrium state with very low
protein density and negligible spatial correlation. Increasing protein
occupancy towards half filling could (and often does) drive the model
system into a striped state with long range order. The
inversion symmetry
we have described dictates that a further increase in density
must eventually destroy striped order. The latter transition may be
more easily envisioned as a consequence of loading thylakoid discs
beyond half filling -- once steric energies have been overcome, the
competition underlying striped order becomes imbalanced, and an
unmodulated state of high density is thermodynamically optimal.
Mathematically, the loss of modulated order at high protein density is
simply the dual transition of its appearance.

Like the lattice gas, our thylakoid stack model can be mapped exactly
onto a spin model with binary variables $\sigma= 2n - 1=\pm 1$. Among
the expansive set of spin models that have been explored numerically
and/or analytically, we are not aware of one that maps precisely onto
this variant of the lattice gas. Many, however, share similar ordering
motifs and spin coupling
patterns. \cite{Ellis2005ar,Deserno1997ar,EzZahraouy2004ar}
Alternating attraction and repulsion in Eq.~\eqref{eq:ham}
correspond to mixed ferromagnetic and antiferromagnetic couplings in a
spin model, \textit{e.g.}, in axial next-nearest neighbor Ising (ANNNI) models,
which can also support modulated order. \cite{FisherSelke1980ar} A
different class of spin models seems better suited to the hard
constraint limit of Eq.~\eqref{eq:ham}. For $\epsilon'=\infty$ each lateral
position on a thylakoid disc can adopt three possible states (both
layers empty, and one or the other layer filled), two of which are
statistically equivalent. The similarity to a three-state Potts model
in an external field is more than superficial. Much of the phase
behavior we identify echoes what is known for that model in three
dimensions, \cite{Wu1982ar} even for finite steric repulsion strengths ($\epsilon'$). 
 
The spirit of our approach echoes many previous efforts to understand
basic physical mechanisms of collective behavior in membrane systems,
from lipid domain formation to correlations among sites pinned by
proteins or substrates. \cite{Speck2010ar,MachtaVeatch2012ar,JhoBrewsterSafranPincus2011ar,WestBrownSchmid2009ar,NajiBrown2007ar,MeilhacDestainville2011ar,PasquaMaibaum2010ar, StachowiakSchmid2012ar,Schmid2016ar,Schick2012ar,PutzelSchick2008ar,Heimburg2007,FrazierAlmeida2007,YethirajWeisshaar2007,SchickKeller2008,MachtaVeatch2011,MitraSethna2018,MeerschaertKelly2015} 
By stripping away most molecular details,
simplified descriptions of phase transitions, such as spin models and
field theories, focus attention on the emergence of dramatic
macroscopic response from a few microscopic ingredients. They also
greatly reduce the computational cost of sampling pertinent
fluctuations, which are simply
inaccessible for biomolecular
systems near phase boundaries when considered in full atomistic
detail. This perspective has even been applied to
lipid ordering in
stacks of membrane
layers, but not in
the context of protein ordering or photosynthesis.
\cite{Tayebi2012ar,Tayebi2013ar,HoshinoKomuraAndelman2015ar,HoshinoKomuraAndelman2017ar}

Here we examine equilibrium structure fluctuations of the lattice
model defined by Eq.~\eqref{eq:ham}, using both computer simulations and
approximate analytical theory. We first describe results of Monte
Carlo sampling, which confirm the stability of a striped phase over a
broad range of temperature and density. We then present mean-field
analysis that sheds light on the nature of symmetry breaking and
relationships with previously studied models.

\begin{table*} 
\centering
\begin{tabular}{c | c} 
    Symbol & Definition \\
    \hline
    $L_x$, $L_y$ & Side length of the lattice in $x$ and $y$ directions \\[0.25cm]  
    $L_{z}$ & Number of thylakoid discs in granum stack \\[0.25cm]
    $z$ & Thylakoid disc index ($1 \leq z \leq L_z $)  \\[0.25cm]
    $\alpha$ & Bottom ($\alpha = 0$) or top ($\alpha = 1$) of a thylakoid disc  \\[0.25cm] 
    $i$ & Location of a lattice site in the $xy$-plane of a disc ($1 \leq  i \leq L_x * L_y $)  \\[0.25cm]
    $n_{\alpha,i}^{(z)}$ & Presence ($n_{\alpha,i}^{(z)} = 1$) or absence ($n_{\alpha,i}^{(z)} = 0$) of proteins at a lattice site
    \\[0.25cm]
    $\beta $  & Inverse temperature $(k_B T)^{-1} $ \\[0.25cm]
    $J $ & Strength of lateral protein attraction ($J>0$) \\[0.25cm]
    $\epsilon $ & Strength of vertical stacking attraction ($\epsilon > 0$) \\[0.25cm]
    $\epsilon' $ & Strength of vertical steric repulsion ($\epsilon' > 0$) \\[0.25cm]
    $\mu $ & Protein chemical potential \\[0.25cm]
    $\bar{n}$ & Average total density (see Eq.~\eqref{eq:ordern}) \\[0.25cm]
    $\Delta n$ & Striping order parameter (see Eq.~\eqref{eq:orderdn}) \\[0.25cm]
    $K$ & Net strength of attraction, relative to thermal energy (see Eq.~\eqref{kdef}) \\[0.25cm]
    $F$ & Free energy (see Eq.~\eqref{eq:freeenergy}) \\[0.25cm]
    $p_{n_1 n_2}$ & Probability of microstate $(n_1,n_2)$ of a two-site cluster 
     \\[0.25cm]
    $\delta$ & Boltzmann factor $e^{- \beta \epsilon'}$ for steric repulsion  \\[0.25cm]
     $K^*$ & Attraction strength at which striping transition becomes discontinuous \\ 

\end{tabular}
\caption{Key physical variables and parameters for the thylakoid lattice model.}
\end{table*}

\section{Methods and Results: Monte Carlo simulations}\label{methodmc}

We used standard Monte Carlo methods to explore the phase behavior of
our thylakoid lattice model. Specifically, we sampled the grand
canonical probability distribution $e^{-\beta H}$ for a periodically
replicated system with $L_x=L_y=10$ and $L_z=24$,
over broad ranges of
temperature and chemical potential. This geometry can accommodate
$L_z/2=12$ copies of the striped motif in the central simulation cell.

Within mean field approximations presented in the next section, the
attractive energy scales $J$ and $\epsilon$ are most important in the
combination $4 J + \epsilon$. We therefore define a parameter
\begin{equation} \label{kdef}
	K \equiv (4 J + \epsilon)/k_B T
\end{equation}
and focus on $\beta \mu$, $K$, and $\epsilon'$ as essential control
variables for this model. The ratio $J/\epsilon$ can also be varied;
but for values of $J/\epsilon$ that are not extreme, this ratio is not
expected to affect qualitative behavior. For simplicity,
we limit attention to results exclusively for
values of $J/\epsilon$ very close to 1/4, for which we
have systematically varied $\beta \mu$, $K$, and $\epsilon'$. A
limited set of simulations with
$J/\epsilon = 0.5$ and 1 
support the ratio $J/\epsilon$ as inessential within the
range studied. 
   
\begin{figure}
\includegraphics[width=0.75\textwidth]{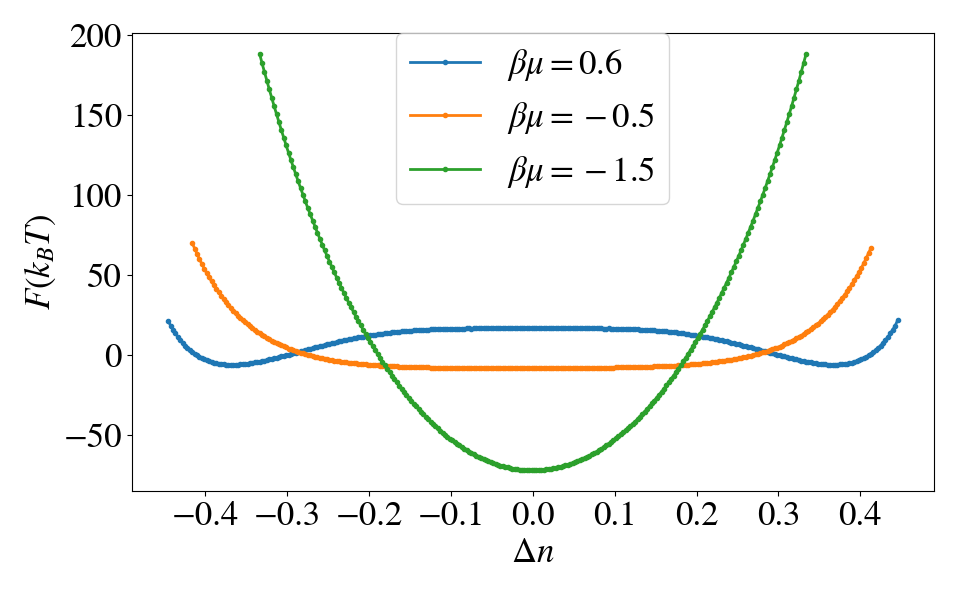}
\caption{\label{fig:mcfree} Statistics of the striping order parameter
  $\Delta n$ at three different thermodynamic states. In all cases
  Monte Carlo simulations were performed with $\epsilon' = 20 k_B T$,
  $J=0.4 k_B T$, $\epsilon = 1.65 k_B T$ (corresponding to $K=3.25$),
  $L_z=24$, and $L_x=L_y=10$. The free energy relative to thermal
  energy, $\beta F = -\ln{P(\Delta n)}$, is shown for $\beta \mu=
  -1.5$, $\beta \mu=-0.5$, and $\beta \mu= 0.6$. For the highest
  value of $\beta\mu$, macroscopic bistability indicates a striped
  state with long-ranged order and broken symmetry. For the lowest
  value of $\beta\mu$, Gaussian fluctuations in $\Delta n$ typify the
  sparse disordered state. For the intermediate value of $\beta\mu$,
  the quartically flat shape of $\beta F$ near $\Delta n=0$ indicates
  proximity to a continuous ordering transition.} 
\end{figure}

These simulations confirm the ordering
scenario described
above, in which the average value $\langle\Delta n\rangle$ of the
striping order parameter can become nonzero in an intermediate range
of $\beta \mu$. In other words, a phase with macroscopically coherent
stripes can be thermodynamically stable at intermediate density. We
identify and characterize transitions between this striped phase and
the ``disordered'' phase with $\langle\Delta n\rangle=0$ by computing
probability distributions $P(\Delta n)$. Fig.~\ref{fig:mcfree} shows corresponding
free energy profiles $F(\Delta n) = -k_{\rm B}T \ln P(\Delta n)$
determined by umbrella sampling (see Supporting Material).
For $2.6 < K < 6$, the progression from
convexity to bistability of $F(\Delta n)$ as $\beta\mu$ increases at
fixed $K$ and $\epsilon'$ is suggestive of Ising-like symmetry
breaking. Quantitative features of $F(\Delta n)$ support this
connection.
In particular, near the transition Binder cumulants approach values
characteristic of 3-dimensional Ising universality (see Supporting Material). 
For $K > 6$ thorough sampling of the equilibrium distribution becomes
challenging, as acceptance probabilities decline due to strong
interactions and striped domains become highly anisotropic. In the Supporting Material we
present indirect evidence that the ordering transition becomes
discontinuous at $K\geq 6$.

Over a wide range of interaction strength $K$,
loading of proteins
into the model thylakoid is thus accompanied by 
continuous transitions in $\langle\Delta n\rangle$, critical
fluctuations, and correspondingly dramatic susceptibility. We locate
this transition through the shape of the free energy profile. The
striped phase is stable wherever $F(\Delta n)$ possesses global minima away
from $\Delta n=0$. Elsewhere, the thylakoid is macroscopically disordered,
though stripe patterns may be prominent on microscopic scales.

\begin{figure*}
\includegraphics[width=0.85\textwidth]{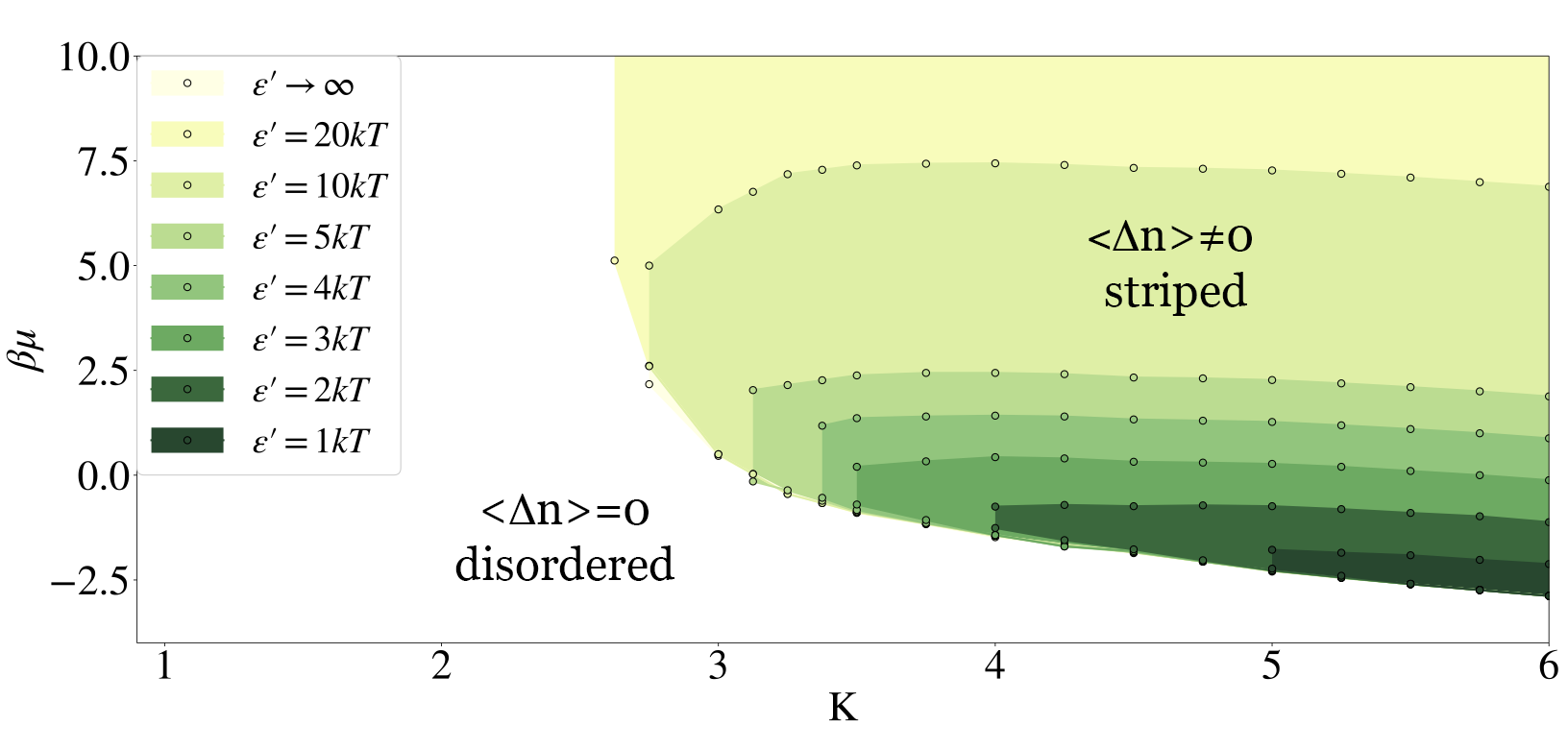}
\caption{\label{fig:phasemc} Phase diagrams of
 the thylakoid lattice model constructed from Monte Carlo simulation
 results, shown in the plane of attraction strength and chemical
 potential. Results are shown for several values of repulsion
 strength $\epsilon'$. In the white region, the disordered phase is
 stable for all $\epsilon'$. The region with darkest shading shows
 the range of $\beta\mu$ and $K$ over which the ordered phase is
 stable for $\beta\epsilon'=1$. The next darkest region shows the
 {\em additional} range of ordered phase stability at
 $\beta\epsilon'=2$, and so on. 
 All phase boundaries, which are assumed to follow
 straight lines between explicitly determined points (circles), mark
 continuous striping transitions.
 Results for the hard constraint limit, $\epsilon'=\infty$,
 are indistinguishable from those with $\beta\epsilon'=20$.
}
\end{figure*}

\begin{figure*}
\includegraphics[width=0.85\textwidth]{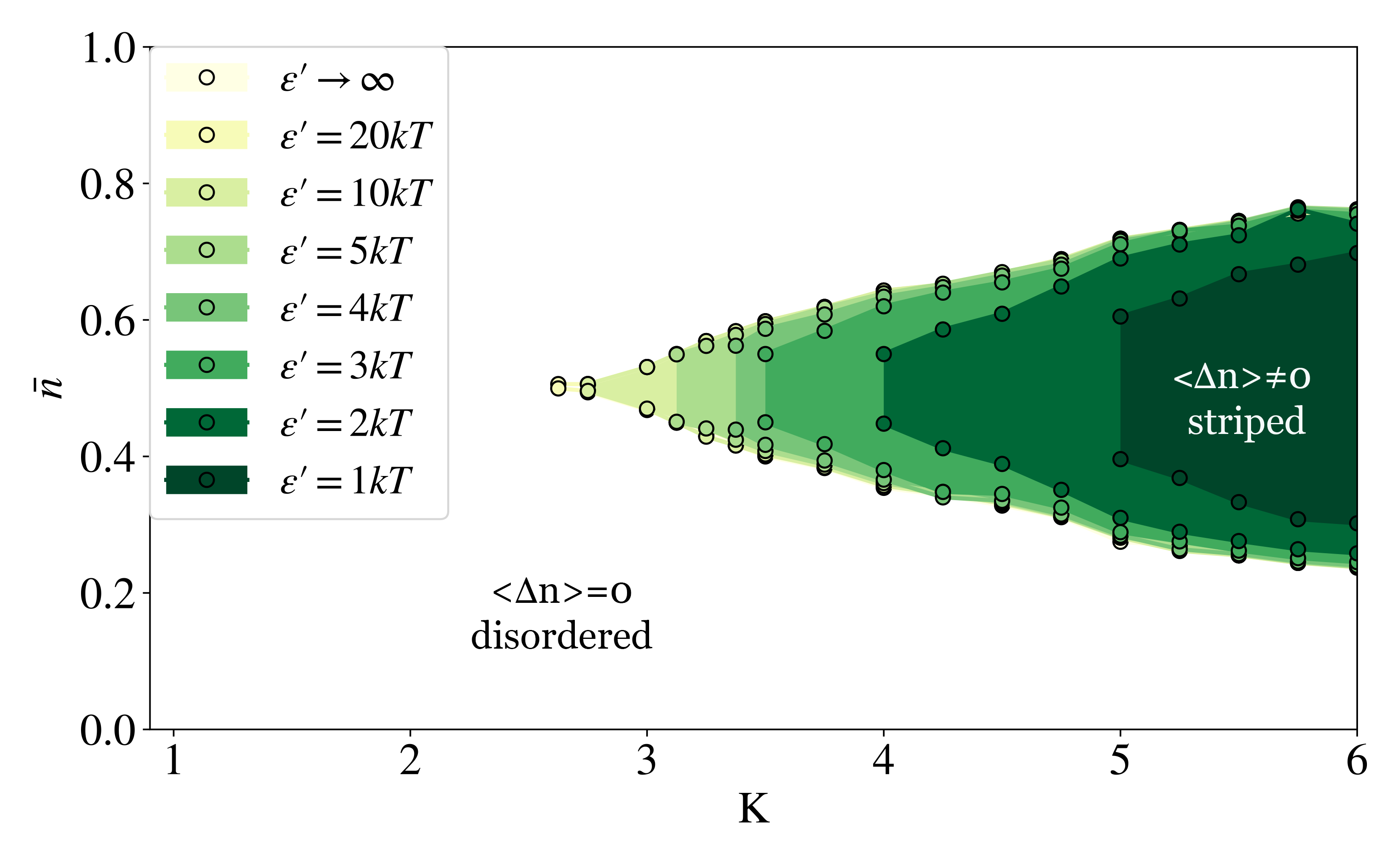}
\caption{\label{fig:phasemcn}Phase diagrams of the thylakoid lattice
 model constructed from Monte Carlo simulation results, shown in the
 plane of attraction strength and density. Points and shading have
 the same meaning as in Fig.~\ref{fig:phasemc} Results for the
 hard constraint limit, $\epsilon'=\infty$, are indistinguishable
 from those with $\beta\epsilon'=20$. For the latter case,
 $\beta\epsilon'=20$, we did not impose high enough chemical
 potential in simulations to obtain results for $\bar{n}>1/2$. In the
 hard constraint limit, the regime $\bar{n}>1/2$ is strictly
 forbidden. } 
\end{figure*}

Fig.~\ref{fig:phasemc} shows the phase diagram in the $(K,\beta\mu)$ plane.
An equivalent but more intuitive representation
in the plane of $K$ and $\bar{n}$
is given in Fig.~\ref{fig:phasemcn}.
 Results are included for a broad range of $\epsilon'$ values. 
 In all cases, computed phase boundaries
are lines of Ising-like critical points. All boundaries
are mirrored across the lines of 
inversion
symmetry of Eq.~\eqref{eq:map}, or 
 $\bar{n}=1/2$ in the $\bar{n}$ vs. $K$ plane, respectively. As described in
Sec.~\ref{mathmodel}, striping transitions 
at finite $\epsilon'$
are
consequently
re-entrant for all finite steric repulsion strengths $\epsilon'$. Modulated
order requires sufficient filling of the lattice but is inevitably
destroyed by high density.

The shapes of these phase diagrams clearly reflect the origin of
modulated order in an interplay between proteins' attraction and
steric repulsion. The domain of stability of the striped phase is
largest where attraction and repulsion are both potent (\textit{i.e.},
$\beta\epsilon'$ and $K$ are both much greater than unity). Small
values of either $\beta\epsilon'$ or $K$ greatly compromise this
stability, or eliminate it entirely.

\section{Methods and Results: Mean field theory }\label{methodmft}

As with most critical phenomena, the long-ranged correlation of
protein density fluctuations implied by these phase transitions
greatly hinders accurate analytical treatment. Here we employ the most
straightforward of traditional approaches for predicting phase
behavior, namely mean field (MF) approximations, to further explore and
explain the ordering behavior revealed by Monte Carlo simulations of
the thylakoid model. Though quantitatively unreliable in general,
mean-field methods provide a simple accounting for the collective
consequences of local interactions, and thus a transparent view of
phase transitions that result.

Mean field theories generically treat the fluctuations of select
degrees of freedom explicitly, regarding all others as a static,
averaged environment. We first consider a pair of fluctuating lattice
sites in a self-consistent field, whose continuous transitions can be
easily inferred. We then analyze an extended subsystem of 12 tagged
lattice sites, whose qualitative predictions align with the simpler
treatment. This consistency suggests a robustness of mean-field
predictions for the thylakoid model.

\subsection{Two-site clusters} \label{twocluster}

In order to describe modulated order of the striped phase, a subsystem
for mean field analysis should include representatives from both
layers of a thylakoid disc. Our simplest approximations therefore
focus on a pair of tagged occupation variables, $n_{1,1}^{(1)}$ and
$n_{2,1}^{(1)}$, describing density fluctuations at vertically
neighboring lattice sites that interact directly through steric
repulsion. We will describe mean field analysis for this two-site
cluster first in the simplifying case $\epsilon'\rightarrow\infty$,
\textit{i.e.}, the hard constraint limit. We then consider the more general
case of finite repulsion strength.

\subsubsection{Hard constraint limit} \label{hardconst}
In the limit $\epsilon'\rightarrow\infty$, the microstate
$n_{1,1}^{(1)} = n_{2,1}^{(1)} = 1$
of our two-site cluster is
prohibited. As a result, the mean field free energy $F_{\rm MF}$ can
be written very compactly. We construct $F_{\rm MF}$ from (a) the
Gibbs entropy associated with probabilities of the cluster's three
allowed microstates and (b) the average energy of interaction with a
static environment. In terms of the order parameters $\bar{n}$ and
$\Delta n$, we obtain
\begin{eqnarray} \label{eq:mftfree}
  \frac{2 \beta F_{\rm MF}}{N}
  &&= - 2\beta\mu \bar{n} + (\bar{n} + \Delta n)\log(\bar{n} + \Delta n)  \nonumber \\
		&&+ (\bar{n} - \Delta n)\log(\bar{n} - \Delta n) \nonumber  \nonumber \\
		&&+ (1 - 2 \bar{n} )\log(1 - 2 \bar{n}) - K (\bar{n}^2 + \Delta{n}^2),
\end{eqnarray} 
where $N$ is the total number of lattice sites. Eq.~\eqref{eq:mftfree}  suggests a
close relationship between our thylakoid model and the well-studied
3-state Potts model of interacting spins. Applying the Curie-Weiss MF
approach to that Potts model yields a free energy
of identical form to Eq.~\ref{eq:mftfree}
for the case of an external field that couples symmetrically to two of the
spin states. \cite{Wu1982ar} The MF phase behavior of the two models is therefore
isomorphic, involving both first-order and continuous
symmetry-breaking transitions. The continuous transitions are
qualitatively consistent with results of 
our Monte Carlo sampling. The discontinuous transitions were not observed in thylakoid model 
simulations for $K<6$; evidence for them emerges only for larger values of $K$, where
sampling becomes challenging.

\begin{figure}
\centering
\includegraphics[width=0.85\textwidth]{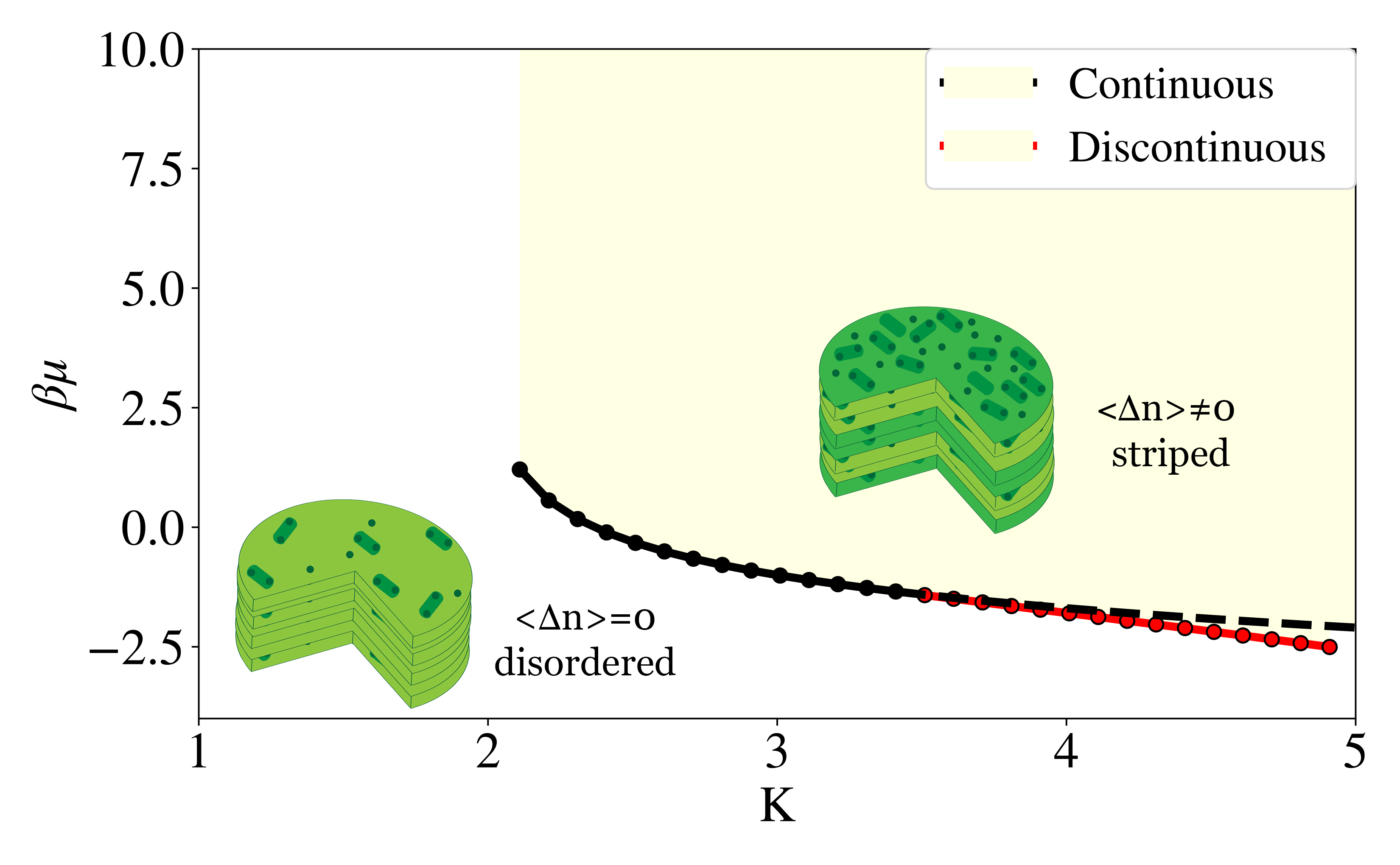}
\caption{\label{fig:mft}  Phase diagram of the thylakoid
  lattice model determined from mean field theory in the hard
  constraint limit $\epsilon'=\infty$, shown in the plane of
  attraction strength and chemical potential. In the white region,
  Eq.~\ref{eq:mftfree} has a single minimum, at $\Delta n=0$,
  indicating a lack of striped order. In the shaded region, global
  minima at nonzero $\Delta n$ indicate symmetry breaking, \textit{i.e.},
  striping with long-range coherence. The extremum of $F_{\rm MF}$ at
  $\Delta n=0$ changes stability at the black curve, allowing for
  continuous ordering. At large $K$ this continuous change is preempted
  by a first-order transition (red curve).}
\end{figure}

 Continuous transitions may be identified by expanding Eq.~\eqref{eq:mftfree} 
 for small $\Delta n$. This expansion indicates a local instability to
symmetry-breaking fluctuations that first appears at
$\bar{n}=K^{-1}$.
A corresponding phase boundary in the $(K,\beta\mu)$ plane can then
be found by minimizing $F_{\rm MF}$ with respect to $\bar{n}$, yielding
$\beta \mu = -1 - \ln(K - 2)$.
 This result, plotted as the black curve in Fig.~\ref{fig:mft}, captures the most basic features of
our simulation results at large $\beta \epsilon'$. As is typically
true, the maximum temperature at which ordering occurs is
overestimated by MF theory (\textit{i.e.}, the minimum value of $K$ is
underestimated).

For sufficiently large $K$, numerical minimization of $F_{\rm MF}$
reveals transitions that are instead discontinuous, as shown
by the red curve in Fig.~\ref{fig:mft}. 
 Here, the disordered state remains locally stable
while global minima emerge at nonzero $\Delta n$. The onset of such
transitions at $K^*=10/3$ can be determined by careful
Taylor expansion of $F_{\rm MF}$ in powers of $\bar{n}$ 
and $\Delta n$ (see Supporting Material). Both of
these order parameters suffer discontinuities at the first-order phase
boundary. For $K < K^*$, no discontinuous transitions are observed
(in Fig.~\ref{fig:mft}, the red curve thus begins at $K^*$).

The absence of first-order transitions in computer simulations
near $K=10/3$
could
signal a failure of this simple mean field theory. Alternatively, such
transitions may occur only at
higher values of $K$.
This low-temperature regime is challenging to explore with
our
Monte Carlo
sampling methods. Below we will show that discontinuous
transitions survive in more sophisticated MF treatments, suggesting
they are a real feature of the model that is difficult to access with
simulations.

Both simulations and MF theory indicate that the striping transition
is not re-entrant in the hard constraint limit. High-density disordered
states are prohibited by steric repulsion at $\epsilon'=\infty$.

\subsubsection{Soft steric repulsion} \label{softconst}
The same basic MF approach can be followed for finite $\epsilon'$. In
this case, however, $F_{\rm MF}$ is written most naturally not as a
function of $\bar{n}$ and $\Delta n$, but instead in terms of
probabilities $p_{n_1 n_2}$ for the four possible cluster microstates:
\begin{align}
  \label{eq:freeenergy}
        \frac{2 \beta F_{\rm MF}}{N}  &= p_{00} \ln p_{00} + p_{10} \ln p_{10} +
        p_{01} \ln p_{01} + p_{11} \ln p_{11}
        \nonumber \\ 
        &- \frac{K}{2} [(p_{11}+p_{10})^2 + (p_{11}+p_{01})^2 ]
        \nonumber \\
        &+ \beta \epsilon' p_{11} - \beta  \mu ( p_{10} + p_{01}  + 2 p_{11} )
\end{align} 
Recognizing that
$\Delta n = (p_{10}-p_{01})/2$ and $\bar{n}=(p_{10}+p_{01}+2p_{11})/2$,
expansion and minimization of Eq.~\eqref{eq:freeenergy} yields 
continuous transitions in the $(K,n)$ plane along
\begin{equation}\label{eq:softmftnans}
  \bar{n} =
\frac{1}{2} \pm \frac{1}{2K}
  \sqrt{(K-2)^2 - 4\delta}
\end{equation}
where $\delta = e^{- \beta \epsilon'}$. The two values of $\bar{n}$
for each $K>2(1+\sqrt{\delta})$ mark transitions to the low- and
high-density disordered phases, reflecting the
occupation inversion symmetry
discussed in
Sec.~\ref{mathmodel}. In the $(K,\beta\mu)$ plane these 
transitions occur at
\begin{equation}\label{eq:softmftaans}
  \beta\mu = \beta\epsilon' - K \bar{n} + \ln{(K \bar{n}-1)}
\end{equation}
where $\bar{n}$ refers to either solution of
Eq.~\ref{eq:softmftnans}. Viewed as functions of $K$ at given 
$\epsilon'$, the two branches of $\beta\mu$ in Eq.~\ref{eq:softmftaans} 
have the peculiar feature of crossing at a certain attraction strength
$K=K_{\rm cross}(\epsilon')$ (see Supporting Material). 
For $K>K_{\rm cross}$
these solutions violate fundamental stability criteria of thermodynamic
equilibrium (see Supporting Material) and therefore cannot be global minima of
the free energy. Lower-lying minima indeed appear at $K^*<K_{\rm cross}$,
preempting the continuous ordering transition before the two solutions
cross.

\begin{figure*}
\includegraphics[width=0.8\textwidth]{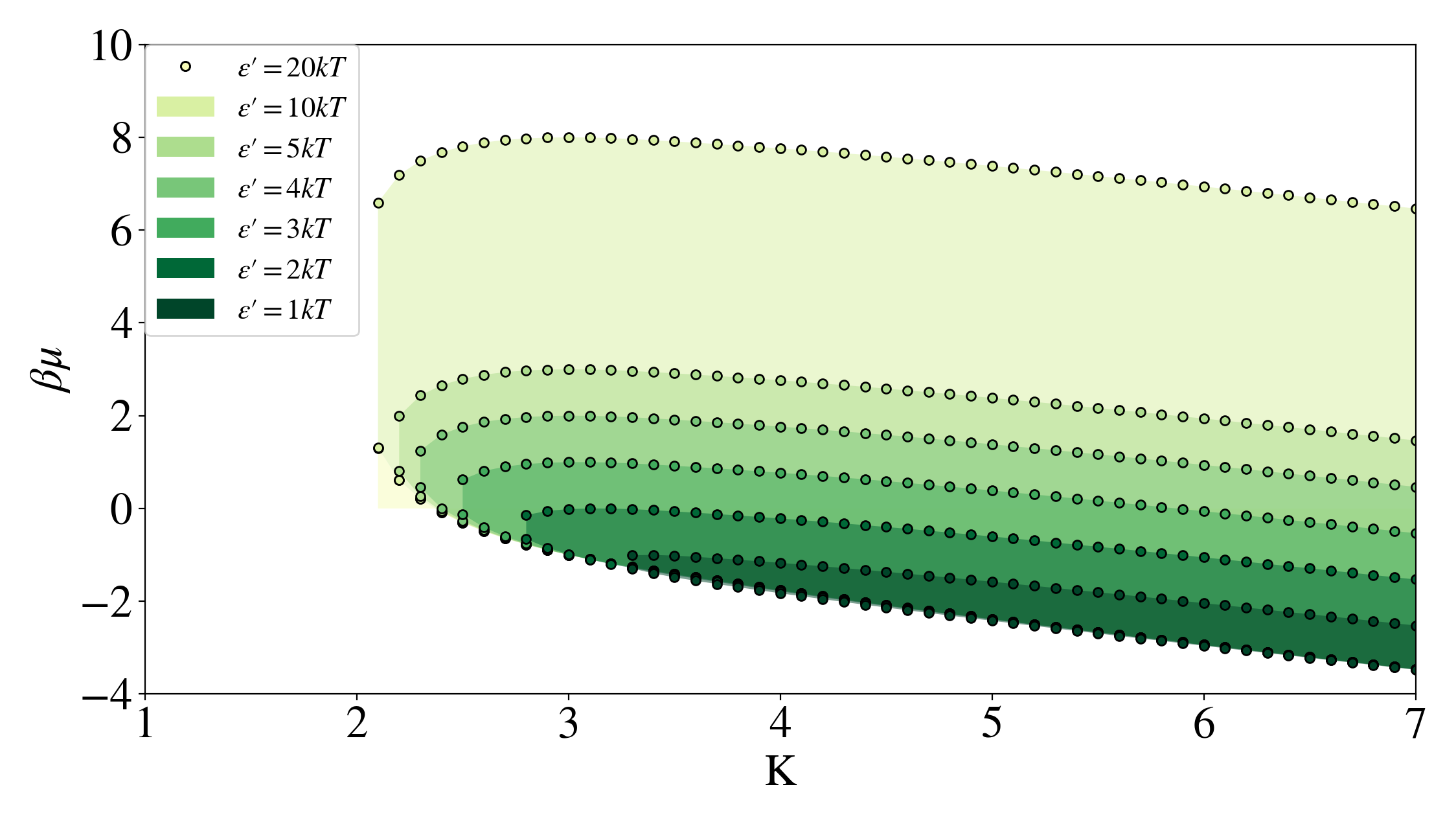}
\caption{\label{fig:musoftmft} Mean-field phase diagram of the
 thylakoid lattice model at finite $\epsilon'$, shown in the
 $(K,\beta\mu)$ plane. Shading has the same meaning as in
 Fig.~\ref{fig:phasemc}. Phase boundaries, determined by minimizing
 Eq.~\ref{eq:freeenergy}, are continuous at small $K$ and
 discontinuous beyond a value $K^*$ that is well approximated by
 Eq.~\ref{eq:firstpredsmall}.}
\end{figure*}

\begin{figure*}
\includegraphics[width=0.8\textwidth]{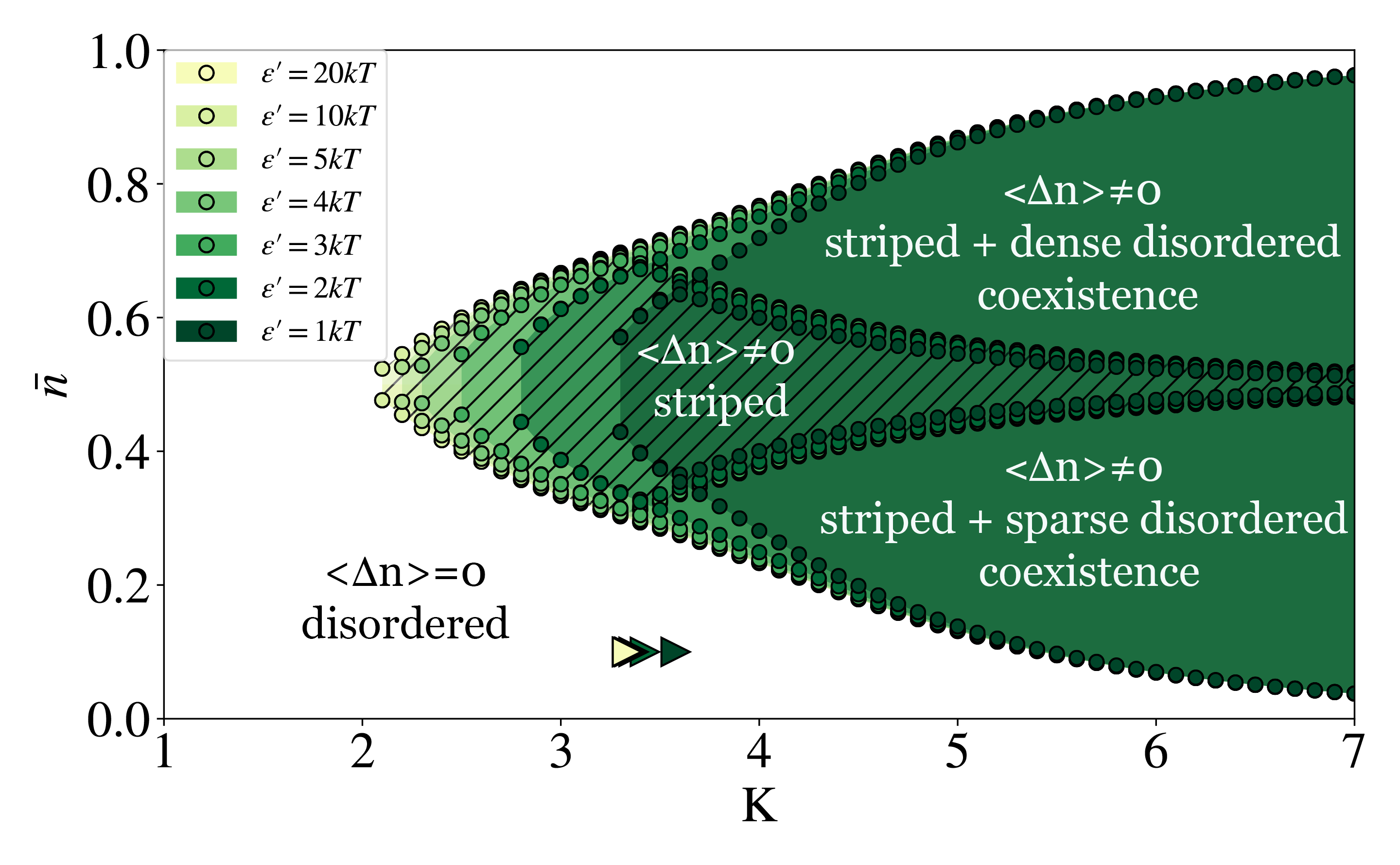}
\caption{\label{fig:nsoftmft}
Mean-field phase diagram of the
 thylakoid lattice model at finite $\epsilon'$, shown in the
 $(K,\bar{n})$ plane. Hatched regions indicate the striped phase, 
 and the coloration corresponds to that of Fig.~\ref{fig:phasemc}. 
 Shaded but un-hatched regions mark coexistence 
 between striped and disordered phases. 
 Phase boundaries, determined by minimizing
 Eq.~\ref{eq:freeenergy}, are continuous at small $K$ and
 discontinuous beyond a value $K^*$ (indicated by triangle markers, with the same color as corresponding $\epsilon'$) that is well approximated by
 Eq.~\ref{eq:firstpredsmall}.
}
\end{figure*}

The development of nonzero $\langle \Delta n\rangle$ with increasing
density
is thus predicted to become discontinuous at sufficiently low temperature, as
in the hard constraint case. The onset of this first-order transition,
\begin{equation}\label{eq:firstpredsmall}
K^* = \frac{10}{3} + \frac{2}{3}\delta + {\cal O}(\delta^2),
\end{equation}
can be determined by Taylor expansion of $F_{\rm MF}$ in the regime of
strong repulsion, \textit{i.e.}, large $\epsilon'$ and small $\delta$.
Figs.~\ref{fig:musoftmft} and \ref{fig:nsoftmft} show mean field phase
diagrams for several values of $\epsilon'$, as determined by numerical
minimization of Eq.~\ref{eq:freeenergy}. For this mean field method, it
is unnecessary to assume a value for $J/\epsilon$, as the mean field
blurs distinctions between vertical and in-plane 
couplings for sites coupled via $J$ or $\epsilon$.
As in the simulation results of Figs.~\ref{fig:phasemc} and \ref{fig:phasemcn}, the data
in Figs.~\ref{fig:musoftmft} and \ref{fig:nsoftmft}
exhibit the symmetry guaranteed by duality.
Discontinuous changes in density upon striping imply regions of
coexistence in the plane of $K$ and $\bar{n}$.
For densities that lie 
between average values for the ordered and disordered states,
both phases are present at equilibrium, as indicated
in Fig.~\ref{fig:nsoftmft}, separated by an interface.

The domain of stability of the striped phase in mean field theory evolves
with $\epsilon'$ in the same basic way observed in Monte Carlo
simulations.
Relative to simulations, however, mean field results are
consistently shifted to
lower $K$ (higher $T$), increasingly so as $\epsilon'$ decreases. 
The discontinuous nature of mean-field transitions at high $K$ is not
easily corroborated by simulations,
as sampling becomes challenging
at high $K$. Limited simulations
with very strong interactions suggest that
first-order transitions appear between $K = 6$ and $K= 7$,
in contrast to the mean field crossover prediction
of $3< K^* < 4$.
Consequently, the coexistence regions
displayed in Fig.~\ref{fig:nsoftmft}
for mean field theory
do not appear in simulation for $K<6$. 

\subsection{Bethe-Peierls approximation}\label{bp}

The accuracy of MF theory is generally improved by examining a larger
set of fluctuating degrees of freedom. \cite{Fair1965ar} In some cases, considering
large clusters can even remove spurious transitions suggested by
lower-level calculations. MF treatments of anisotropic Ising models,
some of which incorrectly predict discontinuous transitions, are
particularly interesting here. Neto \textit{et al.} have surveyed an
array of MF approaches for one such model in two dimensions, which
supports modulated order at low temperature. The simplest MF
calculations predict a crossover from continuous to discontinuous
ordering. The Bethe-Peierls (BP) approximation, a more sophisticated
MF approach, captures the strictly continuous ordering observed in
computer simulations.\cite{Neto2006ar} 

We have performed BP analysis for the thylakoid model (in 3
dimensions), in order to test the robustness of phase behavior
predicted by the two-site calculations described above. Here, we
enumerate all microstates of a subsystem that includes
$n_{1,1}^{(1)}$, $n_{2,1}^{(1)}$, {\em and} all of their remaining
nearest neighbors, a total of 12 sites. The additional sites
experience effective fields representing interactions that are not
explicitly considered. For the specific
case $J=\epsilon$, only two of
these fields may be distinct, greatly simplifying the self-consistent procedure.
We focus exclusively on this case. The calculation and
phase diagrams that result are presented in Supporting Material.

Like simpler MF approaches, the BP approximation yields several
solutions for the effective fields at low temperature. Some of these
solutions correspond to continuous ordering transitions, which can
also be identified by Taylor expansion of the self-consistent
equations. Other solutions describe symmetry-broken states that do not
appear continuously, resembling in many respects the first-order
transitions predicted by two-site calculations. Demonstrating that
these states are thermodynamic ground states would require formulating
a free energy for this BP approach, which we have not pursued. Their
local stability, however, is clearly preserved in the BP scheme.

The most pronounced difference between BP phase diagrams and those of
simpler MF treatments is a shift of phase boundaries to lower
temperature (higher $K$). Agreement with Monte Carlo simulations is
therefore improved. With this shift, the onset of discontinuous
ordering transitions suggested by BP calculations occurs near $K=6$.
This result supports
the notion that first-order transitions are a real feature of the
thylakoid model, occurring near the temperature range suggested by
flat histogram sampling; see Supporting Material.

\section{Discussion}\label{discuss}

The model we have constructed to study vertical arrangement of
proteins in grana stacks is deliberately sparse in microscopic
detail. It does not distinguish among the associating protein species
in photosynthetic membranes, nor does it account for shape
fluctuations of lipid bilayers in which these proteins reside. These
unresolved features are undoubtedly important for the physiological
consequences of the ordering scenario we have described, but they are
inessential to its origins.  The finite size of grana stacks is also
neglected here.  Sharp transitions we have described would
be rounded in real thylakoids by finite size effects, but natural
photosynthetic membranes should be large enough to exhibit
micron-scale cooperativity in protein rearrangements.

A model that distinguishes among LHCII, PSII, and their
super-complexes, we assert, should exhibit the same basic phase
behavior as our coarse-grained representation, so long as it
incorporates the same fundamental competition between attraction and
repulsion. But such a model would be much less amenable to the
theoretical and computational analyses we have presented, particularly
if it acknowledges the differences in these proteins' sizes and
shapes.  PSII proteins are typically more oblong and significantly
larger than LHCII proteins; these two species also extend from the
membrane layer to different degrees, influencing their interactions
with proteins on both adjacent sides. Capturing these geometric
features would greatly complicate a lattice representation. It would
also require commensurately detailed parameterization of
protein-protein interactions; given the scarcity of experimental
information, locating physiologically relevant regions of parameter
space would be even more challenging than for our lattice
description. In conducting such a broad survey with a computationally
demanding model, the existence and stability of an ordered phase could
easily escape discovery altogether.

The lattice model of Eq.~\eqref{eq:ham} is thus well-suited to reveal
the existence and basic nature of vertical ordering in a system with
the geometry and varied interactions of a stack of thylakoid
membranes, which is the focus of this paper.  Confidently establishing
the conditions at which ordering occurs, and assessing its functional
implications, calls for models with greater microscopic detail, like
that of
Ref.\cite{SchneiderGeissler2013ar,SchneiderGeissler2014ar,Schneider2013ar}.
In particular, distinguishing among LHCII, PSII, and super-complexed
PSII-LHCII would enable correlation of vertical ordering with protein
speciation. It would similarly help to explore connections between
vertical ordering and the emergence of horizontally extended arrays of
PSII-LHCII supercomplexes. Previous work suggests that such lateral
ordering can be significantly stabilized by stacking attractions
between thylakoid layers whose PSIIs are appropriately aligned
\cite{StandfussKuhlbrandt2005ar,Daum2010,NieldBarber2000,Boekema2000},
providing a natural mechanism of communication between these modes of
organization.

In models that distinguish among different protein species,
a thylakoid's protein composition, and the LHCII:PSII ratio in
particular, will figure importantly in any ordering scenario.
If, maintaining the same net protein density, the
concentration of LHCII were increased uniformly
throughout
the central region of a granum stack, the balance between steric forces (involving PSII) and stacking
attraction (involving LHCII) would shift. 
In the language of our lattice model, the effective value of $K$ would increase, and $\epsilon'$ would proportionally
decrease.
By looking at Figs.~\ref{fig:phasemcn} and ~\ref{fig:nsoftmft}, one can see that in the density range $0.5 \leq \bar{n} \leq 0.8$, this compositional shift
would favor either the striped
phase, or a coexistence of striped and disordered dense phase, depending on the magnitude of $K$. For
densities greater than 0.8 and sufficiently large $K$, the system would enter
 the coexistence of striped and dense disordered phase.

Our model predicts that, at moderate to high protein density, LHCII
stacking and PSII volume exclusion can induce coexistence between
vertically striped and uniform phases in a stack of many membrane
layers.  Interestingly, previous work suggests that high density
favors a coexistence scenario for \textit{lateral} ordering as well.
According to both experimental and computational studies, the extent
of lateral ordering is quite sensitive to protein packing fraction.
High density of LHCII in particular is known to destabilize lateral
ordering. \cite{SchneiderGeissler2013ar,SchneiderGeissler2014ar,Schneider2013ar,Jansson1997,Veerman2007,KirchhoffAlbertsson2007,Murphy1986,KirchhoffTremmel2004,Haferkamp2010,Kouril2013}
Computational studies suggest that, under such circumstances ($0.65
\leq \bar{n} \leq 0.85$, with LHCII:PSII at most 5:1), crystalline
ordered phases should appear only in coexistence with fluid
phases.\cite{SchneiderGeissler2013ar,SchneiderGeissler2014ar,Schneider2013ar}
Physiologically relevant packing fractions can be similarly high (in
the range 0.6-0.8), and measured LHCII:PSII ratios are in the range
2-6\cite{Jansson1997,Veerman2007,KirchhoffAlbertsson2007,Murphy1986,KirchhoffTremmel2004,Haferkamp2010,Kouril2013},
suggesting that coexistence between crystalline order and laterally
disordered phases predominates both \textit{in vitro} and \textit{in
  vivo}. A tendency for phase separation in the context of both
vertical and lateral ordering suggests that even weak coupling
between the two could induce strong correlation.

 The biological relevance of
structural
rearrangements predicted by our thylakoid model
depends on the
effective physiological values of parameters like $K$, $\epsilon'$,
and $\beta\mu$. Inherent weakness of attraction or repulsion, or else
extreme values of protein density, could prevent thylakoids from
adopting a striped phase. Photosynthetic membranes, however, visit
states in the course of normal function that vary widely in protein
density and in features that control interaction strength. We
therefore expect significant excursions in the
parameter space of Figs.~\ref{fig:phasemc} and \ref{fig:phasemcn}.
Since ordering transitions in our model require only
modest density and interactions not much stronger than thermal energy,
we expect proximity to phase boundaries to be likely in natural
systems. Biological relevance depends also on the functional
consequences of striped order. Photochemical kinetics and
thermodynamics are determined by details of microscopic structure that
we have made no attempt to represent, in particular, gradients in $pH$.
If those aspects of intramolecular and supermolecular
molecular structure are sensitive to local
protein density or to the nanoscale spacing between dense regions, then
striping transitions could provide a way to switch sharply between
distinct functional states.

Given the limited availability of thermodynamic measurements on
photosynthetic membranes, making quantitative estimates of the control
variables $K$, $\epsilon'$, and $\beta\mu$ for real systems is very
challenging. We will focus on the current qualitative knowledge of
properties that are conjugate to these parameters, in order to explore
which phases could be pertinent to which functional states.

The majority of precise measurements on grana have assessed the
density of specific proteins, which is of
course conjugate to their chemical potential. For this reason we have
presented phase diagrams in terms of both $\beta\mu$ and $\bar{n}$. Densities in central grana tend to be quite high, relative to other
biological membranes.  Combining information about the number
density of PSII (from AFM or EM images) with measured number
densities or stoichiometries of other protein species relative to PSII
(typically measured by gel electrophoresis), packing fractions
approach $\bar{n} \approx 0.6-0.8$.
\cite{Schneider2013ar,Kouril2013,Haferkamp2010,KirchhoffTremmel2004,Murphy1986}
Therefore, the top half of Figs.~\ref{fig:phasemcn} and
~\ref{fig:nsoftmft} are likely the most relevant for biology.

The net attraction strength relative to temperature, $K$, is conjugate
to the extent of protein association within each membrane layer and
across the stromal gap. Because experiments suggest stacking 
interactions have an empirically measured,
dramatic effect on 
protein association, \cite{ChowKimHortonAnderson2005ar,StandfussKuhlbrandt2005ar,PhuthongHuangGrossman2015ar,OnoaSchneider2014ar,RubanJohnson2015ar} we will focus on the extent of stacking
as a rough proxy for $K$. 
Previous computational work suggests that the range of $K$ we have explored is
physiologically reasonable. Focusing on lateral protein ordering in a
pair of membrane layers,
Refs. \cite{SchneiderGeissler2013ar,Schneider2013ar} found that
configurations consistent with atomic force microscopy images could be
obtained for weak in-plane protein-protein attractions of energy $\leq
2 k_BT$ and stacking energy $4 k_BT$. Associating the
energy scales of that particle model with the energies of our more
coarse-grained lattice representation ($\beta J \lesssim 2 $ and $\beta
\epsilon
\approx 4 $) suggests values of $K$ in the neighborhood of
5-10. 
Ordering at moderate to large $K$ is therefore likely to be most
physiologically relevant.  It is in this regime that we find a
evidence for a crossover in the nature of the striping transition, from
continuous to discontinuous.

The strength of steric repulsion, $\epsilon'$, is strongly influenced
by thylakoid geometry. For a very narrow lumen and very rigid
lipid bilayers, PSII molecules on opposite sides of a thylakoid
disc are essentially forbidden to occupy the same lateral position, a
hard constraint that is mimicked by the limit $\epsilon'=\infty$ of
Sec.~\ref{hardconst}. Greater luminal spacing, together with membrane flexibility,
abates or possibly nullifies this repulsion. We therefore regard
thylakoid width as a rough readout of $\epsilon'$. 
Since thylakoid width changes significantly as light conditions
change, we also view $\epsilon'$ as a control variable related to
light intensity. 
Each thylakoid is roughly 10-15 nm
thick, and the lumenal gap measures approximately 2-7 nm, depending on light conditions. \cite{DekkerBoekema2005ar,PribilLabsLeister2014ar,NevoCharuviTsabariReich2012ar}

In high light conditions, the luminal gap of the thylakoid discs
widens. \cite{PuthiyaveetilTsabari2014ar,KirchhoffDelbruck2012ar}
This geometric change should ease steric repulsion, though lumen 
widening is less substantial at the center of the discs than at their edges.
\cite{ClausenBrooksNiyogiFletcher2014ar,IwaiPackTakenakaSakoNakano2013ar}
If the light
intensity is particularly high, this expansion
can be accompanied by the disassembly of
PSII-LHCII mega-complexes (and, to a much lesser extent, super-complexes)
en route to PSII repair.
\cite{EricksonWakaoNiyogi2015ar,PuthiyaveetilTsabari2014ar,KirchhoffDelbruck2012ar,HerbstovaKirchhoff2012ar,Koochak2019}
Although this disassembly is primarily
limited to the edges of the thylakoid, we infer an overall decrease in
the extent of stacking. And because PSII is subsequently shuttled to
the stroma for repair, we also expect a concomitant decrease in protein
density.
The implied low to modest values
of $\beta\mu$, $\epsilon'$, and $K$
suggest that high light scenarios favor the sparse disordered
phase of our model.
    
In low light conditions, thylakoid discs are thinner, and the stromal
gaps between them decrease as well \cite{DekkerBoekema2005ar,PribilLabsLeister2014ar}, pointing to large values of $\epsilon'$ and $K$. 
 The low-light state thus appears to be
the strongest candidate for the striped phase we have described.

During state transitions, a collection of changes causes the balance
of electronic excitations to shift from PSII to photosystem
I. \cite{EricksonWakaoNiyogi2015ar,ClausenBrooksNiyogiFletcher2014ar,IwaiPackTakenakaSakoNakano2013ar,Minagawa2011ar,LemeilleRochaix2010ar}.
Among these changes, a diminution of stacking and a shift of LHCII
density towards the stroma lamellae are closely related to the
ordering behavior of our thylakoid model. Both result from
phosphorylation of some fraction of the LHCII population, which
weakens attraction between discs, prompts disassembly of a fraction of
PSII-LHCII mega-complexes and super-complexes, and allows LHCII
migration towards the thylakoid margins. The corresponding reduction
of $\beta \mu$ and $K$ is likely to be highly organism-dependent,
since the extent of phosphorylation varies greatly from algae to
higher plants. \cite{IwaiPackTakenakaSakoNakano2013ar,Minagawa2011ar,LemeilleRochaix2010ar, WlodarczykCroce2016ar,CrepinCaffarri2015ar,Nawrocki2016ar}
Lacking as well quantitative information
about thylakoid thickness, it is especially difficult to correlate state
transitions with the phase behavior of our model. In the case of very
limited phosphorylation (as in higher plants), the ordered and sparse
disordered phases both seem plausible. With extensive phosphorylation
(as in algae), substantial reductions in stacking attraction and
density make the ordered state unlikely.
 
The relationship among granum geometry, protein repulsion strength,
and long-range stripe order suggests interesting opportunities for
manipulating the structure and function of thylakoid membranes in vivo.
By adjusting the luminal spacing, mechanical force applied to a
stack of discs in the vertical direction (\textit{i.e.}, the direction of
stacking) should serve as a handle on the steric interaction energy
$\epsilon'$. The phase behavior of our model suggests that smooth
changes in force can induce very sharp changes in density, protein
patterning, and stack
height. Ref. \cite{ClausenBrooksNiyogiFletcher2014ar} demonstrates a
capability to manipulate thylakoids in this way, and could serve as
a platform for testing the realism of our lattice model. 
Complementary changes in attraction strength might be achieved by
controlling salt concentration, a strategy used in
Ref. \cite{KirchhoffRogner2007ar} to examine the influence of stacking
interactions on lateral ordering of proteins in a pair of thylakoid discs.

\section{Conclusion}\label{conclusion}

The computer simulations and analysis we have presented establish that
ordered stripes of protein density, coherently modulated from the
bottom to the top of a granum stack, can arise from a very basic and
plausible set of ingredients. Most important is the alternation of
attraction and repulsion in the vertical direction, a feature that is
strongly suggested by the geometry of thylakoid membranes. Provided
the scales of these competing interactions are both substantial, a
striped state with long-range order will dominate at moderate density.
Under
conditions accessible by computer simulation, the striping transition
is continuous, with critical scaling equivalent to an Ising model or
standard lattice gas. Mean-field analysis suggests that the transition
becomes first-order for strong attraction, switching sharply between
macroscopic states but lacking the macroscopic fluctuations of a
system near criticality.

Simple mechanisms for highly cooperative switching have been proposed
and exploited in many biophysical contexts, \cite{StachowiakSchmid2012ar,Schmid2016ar} including the lateral
arrangement of proteins in photosynthetic membranes. \cite{NevoCharuviTsabariReich2012ar,LiguoriPerioleMarrinkCroce2015ar,KirchhoffRogner2007ar,EricksonWakaoNiyogi2015ar,PuthiyaveetilTsabari2014ar,ClausenBrooksNiyogiFletcher2014ar,HerbstovaKirchhoff2012ar,Minagawa2011ar,LemeilleRochaix2010ar,WlodarczykCroce2016ar,Nawrocki2016ar} We suggest that
vertical ordering in stacks of such membranes can be a complementary
mode of collective rearrangement with important functional
consequences.

\section{Author Contributions}

A.M.R. performed all of the Monte Carlo simulations except the flat histogram sampling, numerical
solutions, data analysis, and figure generation, as well as developed
all the necessary software. P.L.G. provided guidance in these
tasks and performed the flat histogram sampling in the Supporting Material. A.M.R. and P.L.G. authored this manuscript.

\section{acknowledgments}
We acknowledge the financial support of the National Science Foundation GRFP program, the Hellman Foundation, and National Science Foundation grant MCB-1616982. We thank Anna Schneider for her coarse-grained model of lateral protein organization and its associated code base, which was used to initially explore a model higher plant photosynthetic system. We also greatly appreciate conversations with Helmut Kirchhoff and the groups of Krishna Niyogi and Graham Fleming.


\section*{Supporting Citations}

References \cite{Ruder2017ar,Ferrenberg2018ar,KumarRosenberg1992ar,Binder1981ar,Binder1981Bar,Wang2000ar} appear in the Supporting Material.

\section*{Citations}
\bibliography{rosnik_lattice}

\section*{Supporting Material} \label{si}

\section{Methods and Results: Monte Carlo} \label{simc}

\subsection{Simulation specifications} \label{simmeth}

Phase transitions were determined via umbrella sampling, a form of
biased MC simulations. The bias added to the Hamiltonian energy was a
harmonic potential $\frac{1}{2}k (\langle \Delta n \rangle - \Delta
n_{target})^2$ with a spring constant $k$ of 10,000 $k_B
T$. Simulations were run for (2 to) 3 million MC sweeps, saving
$\Delta n$ and $\bar{n}$ data every
100 sweeps. The bias targets ranged from $\Delta n_{target} = -0.5 $
to $\Delta n_{target} = 0.5 $ for a total of 51 distinct $\Delta
n_{target}$ values. With these data, free energy profiles were
constructed via the WHAM method. \cite{KumarRosenberg1992ar}

\subsection{Binder cumulants } \label{sibind}
\begin{figure}
\centering
\includegraphics[width=0.75\textwidth]{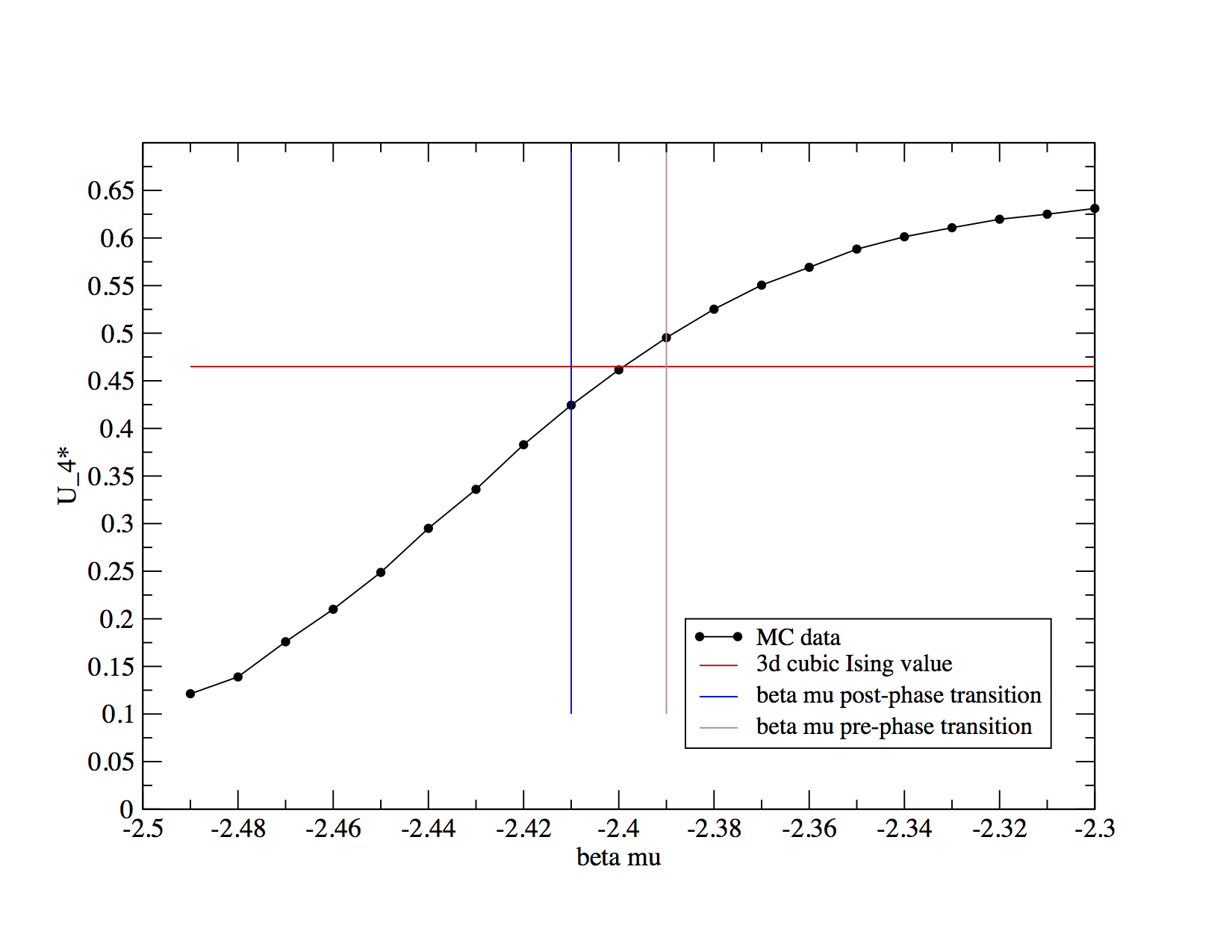}
\caption{\label{fig:bind1low}
 Binder cumulant $U_4^*$ as a function $\beta \mu$ for $J = 0.675 k_B
 T$, $\epsilon = 2.55 k_B T$, and $\epsilon' = 1 k_B T$. The
 horizontal dashed line represents the three-dimensional cubic Ising
 universality value of 0.465 The horizontal red line indicates the
 universal value $U_4^*=0.465$ corresponding to the three-dimensional
 Ising model on a cubic lattice. Vertical lines bracket the range of
 $\beta \mu$ over which $F(\Delta n)$ changes convexity. }
\end{figure}

We computed Binder cumulants for the thylakoid striping transition in
order to verify its Ising universality classification. We specifically
consider \cite{Binder1981ar,Binder1981Bar}
\begin{equation} \label{eq:binder}
 U_4^* = 1 - \frac{ \langle (\Delta n)^4 \rangle } { 3 \langle (\Delta n)^2 \rangle^2 }
\end{equation}
Fig.~\ref{fig:bind1low} shows $U_4^*$ as a function of $\beta\mu$ for
$K=5.25$ and $\beta\epsilon'=1$, over a range that spans the
ordering transition. The interval in which the free energy $F(\Delta
n)$ changes convexity is also marked.
Values of $U_4^*$ in this interval lie near that
expected for the three-dimensional cubic
Ising model universality class. \cite{Ferrenberg2018ar}

\begin{figure}
\centering
\includegraphics[width=0.75\textwidth]{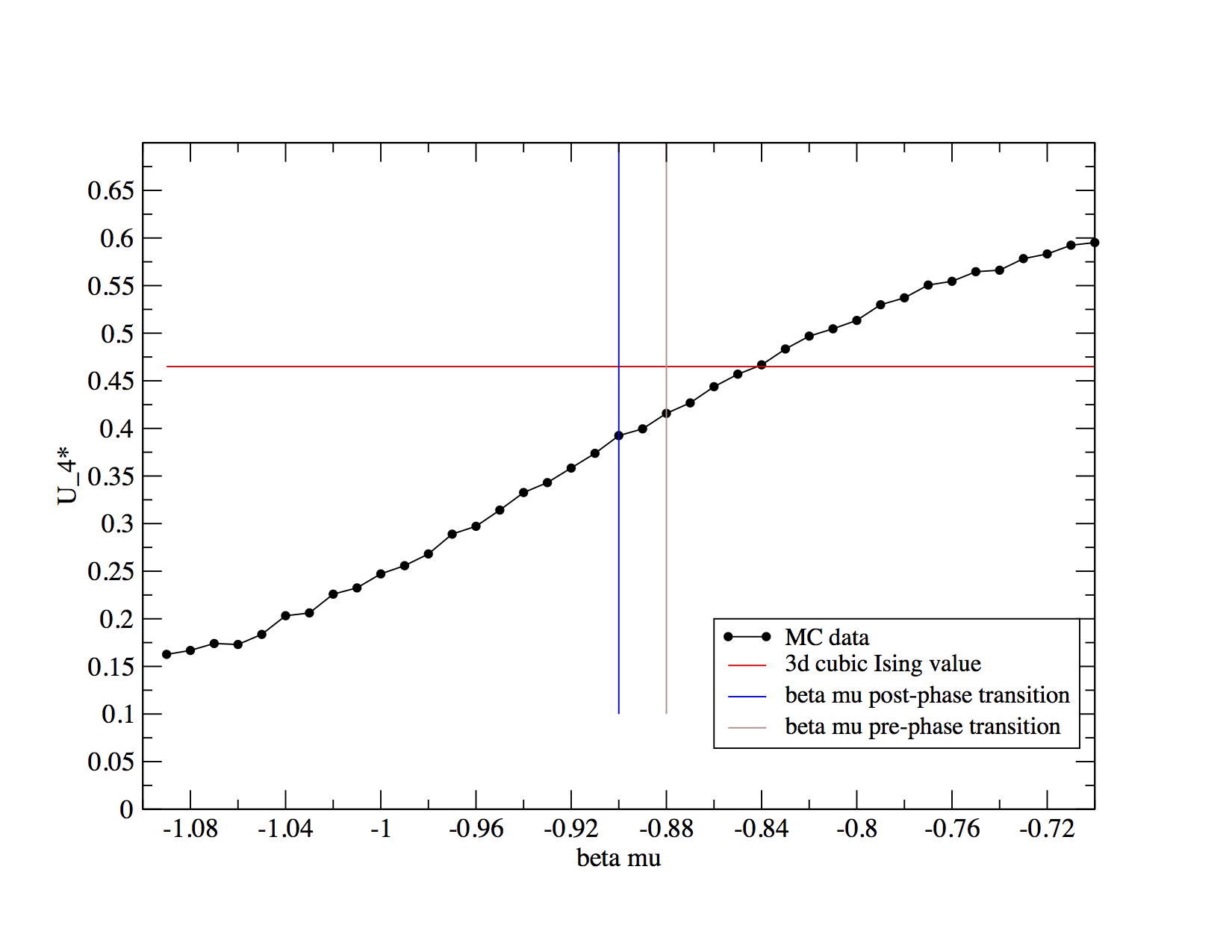}
\caption{\label{fig:bind20} Binder cumulants for $\beta \mu$ at and near transition for $J = 0.45 k_B T$, $\epsilon = 1.7 k_B T$, and $\epsilon' = 20 k_B T$. The horizontal dashed line represents the three-dimensional cubic Ising universality value of 0.465. Vertical lines bracket the range of $\beta \mu$ over which $F(\Delta n)$ changes convexity.}
\end{figure}

Fig.~\ref{fig:bind20} shows analogous results for $K=3.5$ and $\beta\epsilon'=20$.
                        
\subsection{Evidence for first-order transitions in simulation }\label{simcfirst}
Statistics of the order parameters $\Delta n$ and $\bar{n}$ can
be obtained efficiently by routine umbrella sampling only for
interaction strengths below $K\approx 6$. In this range we observe
only continuous ordering in the thylakoid model. In order to evaluate
the mean-field prediction of first-order transitions at high $K$, we
employed a flat histogram sampling method analogous to
Ref.~\cite{Wang2000ar}. Adaptive biasing was applied to a variable $p_{11}$
that couples strongly to the high-density transition. Specifically,
$$
p_{11} = \frac{2}{ L_x L_y L_z} \sum_{z,i}n^{z}_{1,i} n^{z}_{2,i}
$$
quantifies the instantaneous steric repulsion due to protein occupancy
on both sides of a thylakoid disc.
These simulations were performed by PLG.

\begin{figure*}
\centering
\includegraphics[width=0.85\textwidth]{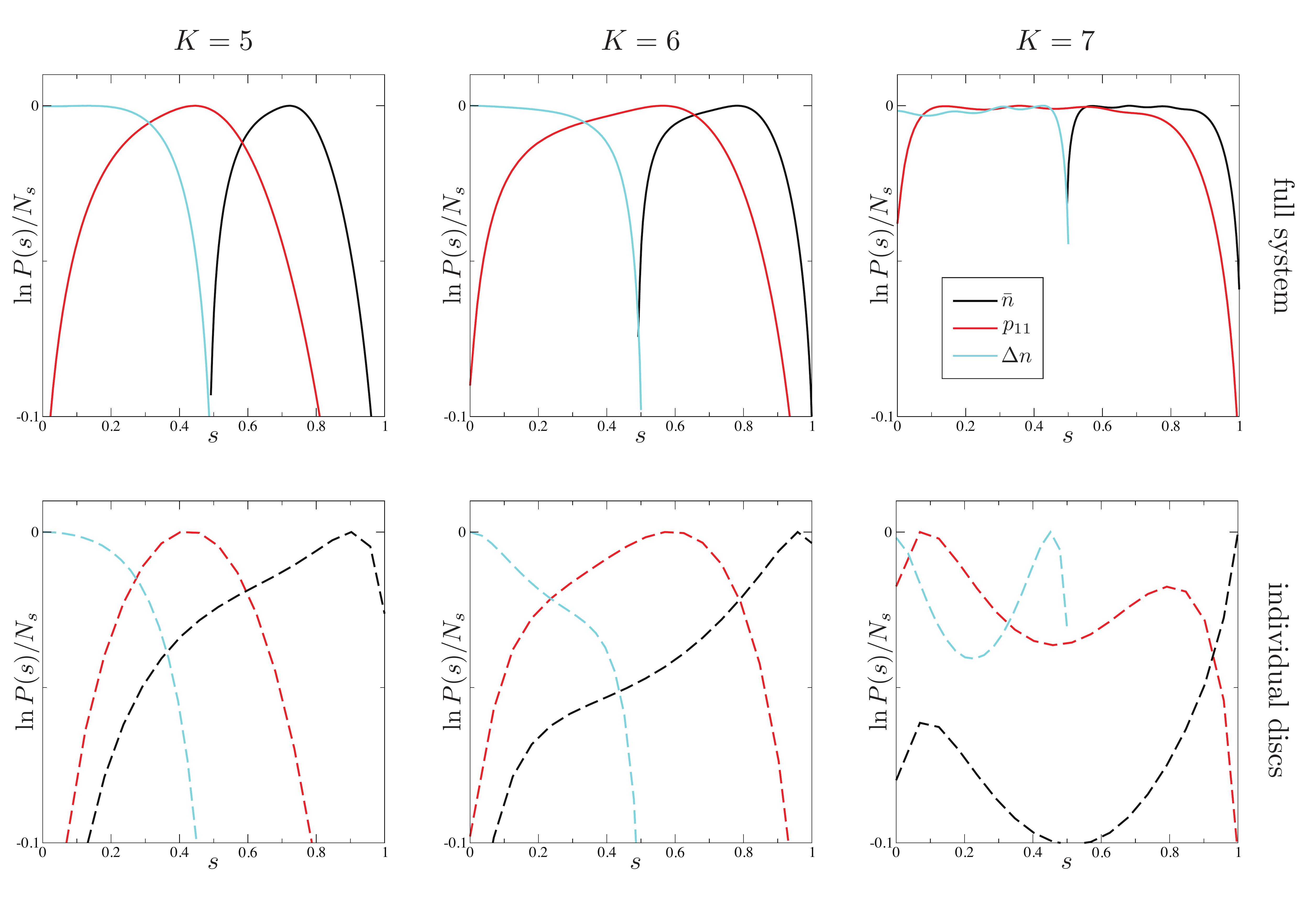} \label{plgsi}
\caption{ Log probability distributions for order parameters
  $\bar{n}$, $\Delta n$, and $p_{11}$ (top row), as well as their
  disc-wise analogs (bottom row).  Fat tails at the ordering
  transition develop as $K$ is increased (moving from left to right in
  the figure columns). Clear multiple peaks at large $K$ strongly
  suggest the macroscopic bimodality underlying discontinuous phase
  transitions.  }
\end{figure*}

Results of this flat histogram sampling are shown in Fig.~\ref{plgsi}
for systems with $L_x=L_y=6$ and $L_z=12$ at three different high
values of $K$, and $\beta\epsilon'=5.5$. Scaled log probabilities are shown for the global order
parameters $\Delta n$, $\bar{n}$, and $p_{11}$ (top panels), and also for their
disc-wise analogs (bottom panels), e.g.,
$$
\bar{n}^{(\rm individual)} = \frac{1}{L_x L_y} \sum_{\alpha,i} n^{z}_{\alpha,i},
$$
where $z$ could refer to any of the discs. (Because discs are
statistically equivalent, we accumulate statistics over all values of
$z$.) The index $s$ specifies one of these six order parameters. For
each $s$, the corresponding scaling factor $N_s$ is chosen so that the
plotted quantities serve as large deviation rate functions: For
$\bar{n}$, $N_s=L_x L_y L_z$; for $\Delta n$, $N_s=L_x L_y L_z/2$; and
for $p_{11}$, $N_s=L_x L_y L_z/2$. For the disc-wise analogs, $N_s=L_x
L_y$ in each case.

For each $K$, we consider a value of $\mu$ that is very
close to the high-density phase boundary, namely $\beta\mu=2.8$ for
$K=5$, $\beta\mu=2.4$ for $K=6$, and $\beta\mu=1.93$ for $K=7$.

For $K=5$, computed distributions are consistent with results of
umbrella sampling described in the main text. Fluctuations of $\Delta
n$ are extremely broad at the transition, and distributions of the remaining
order parameters show no exceptional features.

By contrast, for $K=7$ we observe several features that point towards
discontinuous ordering.  Distributions of extensive parameters acquire
considerable structure, suggesting stiff horizontal domain boundaries
that span the lateral dimensions of a disc.  In this scenario,
appropriate alternation of coexisting striped and doubly occupied
discs can yield very low interfacial free energy, favoring a handful
of specific order parameter values. This same structure, however,
complicates the identification of bistability characteristic of a
first-order transition. Such bistability is instead apparent in the
disc-wise statistics, which are clearly bimodal.

In the intermediate case $K=6$, the statistics of these parameters
show hints of emerging bistability. At the ordering transition each
distribution exhibits fat tails, but none features distinct bimodality.
We therefore estimate the onset of discontinuous ordering somewhere in
the range $6 < K <7$.

\section{Methods and Results: Mean-field theory} \label{simft}
Mean-field phase diagrams were obtained by numerically
minimizing 
the free energy in Eq. (5) or (6)
of the main 
text. We found it most efficient to do so by iterating self-consistent
equations that determine local free energy minima. Here we provide
these self-consistent equations, which result from differentiating
$F_{\rm MF}$, and detail other aspects of our mean-field analysis.

\subsubsection{Self-consistent equations for the hard constraint limit}

	The hard constraint MFT average order parameter is 
			\begin{equation} \label{eq:mftavg}
				\Delta n = \frac{1}{2} \frac{e^{\beta \mu} (e^{K \bar{n}^{(1)}} - e^{K \bar{n}^{(2)}})}{1 + e^{\beta \mu} e^{K \bar{n}^{(1)}} + e^{\beta \mu} e^{K \bar{n}^{(2)}} } 					
			\end{equation}
			where $\bar{n}^{(i)}$ refers to the average density in the $i$th layer. Mutatis mutandis for $\bar{n}^{(2)}$. 
			The average density is
			\begin{equation} \label{eq:eqmft}
				\bar{n} = \frac{1}{2} \frac{e^{\beta \mu} (e^{K \bar{n}^{(1)}} + e^{K \bar{n}^{(2)}})}{1 + e^{\beta \mu} e^{K \bar{n}^{(1)}} + e^{\beta \mu} e^{K \bar{n}^{(2)}} } 	
			\end{equation}

\subsection{Onset of first-order transitions for the hard constraint limit}

We identify the onset of discontinuous transitions by posing the
question: As the free energy extremum at $\bar{n}=1/K$ and $\Delta
n=0$ loses local stability, do lower-lying minima of $F_{\rm MF}$ exist?
Near the onset we assume that such minima reside at very small $\Delta
n$ and at $\bar{n}$ very close to $1/K$; for a given value of $\bar{n}$,
these minima $\Delta n^*$ satisfy
		\begin{equation}\label{eq:k103p3}
			\centering		
                        			\Delta n^{* 2} = 3\bar{n}^3 \bigg(K - \frac{1}{\bar{n}} \bigg)
		\end{equation}
where we have neglected terms of order $\Delta n^4$.

Setting $\bar{n}=1/K+\eta$, Eq.~\ref{eq:k103p3} gives
\begin{equation}\label{eq:k103p4}
			\centering		
			\Delta n^* = \pm\sqrt{ \frac{3}{K} \eta} + O(\eta^{3/2})
\end{equation}

To lowest order in $\eta$, the mean-field free energy
$F_{\rm MF}(\bar{n},\Delta n)$
at the putative
satellite minima can then be written
		\begin{align}\label{eq:k103p5}
			\centering \frac{2\beta}{N} F_{\rm MF}
                        \bigg(1/K + \eta, \pm \sqrt{ \frac{3}{K}
                          \eta}\bigg) &= \frac{2\beta}{N} F_{\rm MF}
                        \bigg( 1/K, 0 \bigg)
                        \nonumber \\ & + \bigg( -3K + \frac{4K}{K-2}
                        \bigg) \eta^2
		\end{align}
                For $K > 10/3$, this free energy lies below that of
                the critical state at $\bar{n}=1/K$ and $\Delta
                n=0$. In other words, symmetry breaking occurs
                discontinuously, before the symmetric state becomes
                permissive of macroscopic fluctuations.
                
\subsection{Self-consistent equations for soft steric repulsion}
Minimizing the free energy Eq. (7) with respect to $p_{10}$, $p_{01}$, $p_{11}$, and
$p_{00} = 1 - (p_{10} + p_{01} + p_{11})$ gives nonlinear expressions
for the mean density in alternating layers,
\begin{equation}
 \label{eq:softmftn1}
 n_1 = p_{10}+p_{11}= \frac{1}{q} ( a e^{K n_1}
 + \delta a^2 e^{K (n_1+n_2)} ) ,
\end{equation}
and
\begin{equation}
 \label{eq:softmftn2}
 n_2 = p_{01}+p_{11}= \frac{1}{q} ( a e^{K n_2}
 + \delta a^2 e^{K (n_1+n_2)} ) ,
\end{equation}
where $a = e^{\beta \mu}$, $\delta = e^{- \beta \epsilon'}$, and
\begin{equation}
q = 1 + a (e^{K n_1}+e^{K n_2}) + \delta a^2
                        e^{K (n_1+n_2)}.
\end{equation}
Iteration of these expressions converges rapidly to local minima of
$F_{\rm MF}$. From these solutions, our primary order parameters are
computed simply from $\bar{n} = (n_1 + n_2)/2$, and $\Delta n = (n_1 -
n_2)/2$.

\subsection{Continuous transitions for soft steric repulsion}
\begin{figure}
\centering
\includegraphics[width=0.75\textwidth]{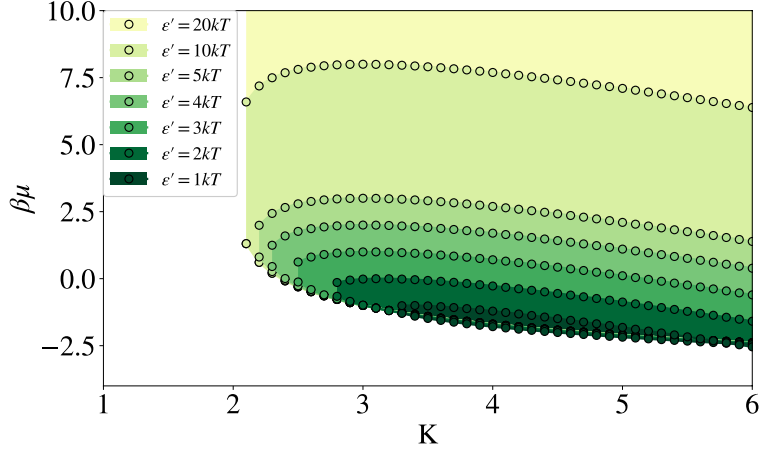}
\caption{\label{fig:softmu} Soft constraint model $\beta \mu$ vs. $K$ phase diagram, continuous mean-field transitions according to Eq.~\eqref{eq:softmftnanssi}. Shaded region indicates the striped phase. }
\end{figure}

For finite $\epsilon'$, the extremum of $F_{\rm MF}$ at $\Delta n=0$
becomes locally unstable when
\begin{equation}\label{eq:softmftnanssi}
  \bar{n} =
      \frac{1}{2} \pm \frac{1}{2K}
      \sqrt{(K-2)^2 - 4\delta},
\end{equation}
defining possible continuous transitions in the $(K,\bar{n})$
plane.
Fig.~\ref{fig:softmu} shows both lines of solutions in the
$(K,\beta\mu)$ plane, for several values of
$\epsilon'$. In each case the two lines cross at an attraction
strength $K_{\rm cross}(\epsilon')$. For $\beta\epsilon'\geq 2$,
$K_{\rm cross}$ lies outside the range of this plot.

Continuous transitions predicted for $K > K_{\rm cross}$ violate a
fundamental thermodynamic requirement of stability. Specifically, the
solution with higher density $\bar{n}$ occurs at a lower chemical
potential than the low-density solution, implying a negative
compressibility. Although these solutions represent local free energy
minima, they cannot be global minima. Indeed, numerical minimization
of $F_{\rm MF}$ identifies lower-lying minima in all cases.
		
\subsection{Self-consistent equations for soft steric repulsion}
Minimizing the mean-field free energy for finite $\epsilon'$
yields nonlinear equations for the average layer densities:
\begin{equation} \label{eq:softmftni}
 \langle n_i \rangle = \frac{1}{q} ( a e^{K n_i} + \delta a^2 e^{2K n} ) ,
\end{equation}
In terms of $n$ and $\Delta n$, 
		\begin{equation} \label{eq:softmftn}
			n = \frac{1}{2q} \bigg[ a e^{K n} ( e^{K \Delta n} + e^{-K \Delta n} ) + 2 a^2 e^{2 K n} \delta \bigg]
		\end{equation}
		\begin{equation} \label{eq:softmftdn}
			\Delta n = \frac{1}{2q} \bigg[ a e^{K n} ( e^{K \Delta n} - e^{-K \Delta n} ) \bigg]
		\end{equation}
		where $a = e^{\beta \mu}$, $\delta = e^{- \beta \epsilon'}$, $n = \frac{1}{2} (n_1 + n_2)$, and $\Delta n = \frac{1}{2} (n_1 - n_2)$. 
		
\subsection{Solving self-consistent equations}
Iterating the self-consistent equations \eqref{eq:softmftn} and
\eqref{eq:softmftdn} converges readily to local extrema of the
mean-field free energy. After $10^6$ steps, additional iteration
changes values of $n_1$ and $n_2$ by less than $10^{-12}$.

Under many conditions, however, this free energy surface exhibits
three or more distinct minima. The end result of iteration thus
depends on initial values of $n_1$ and $n_2$. We considered five
different $(n_1, n_2)$ pairs, namely (0.6, 0.4), (0.1, 0.1), (0.9,
0.9), (0.9, 0.1), and (0.2, 0.1). For each set of conditions, we then
select the self-consistent solution with lowest free energy.

A resulting value of $|n_1-n_2|$ greater than $10^{-9}$ was taken to
signify thermodynamic stability of the ordered phase. 

\section{Methods and Results: Bethe-Peierls approximation}

\subsection{One-cluster expressions}

\begin{figure}[h]
\centering
\includegraphics[width=0.4\textwidth]{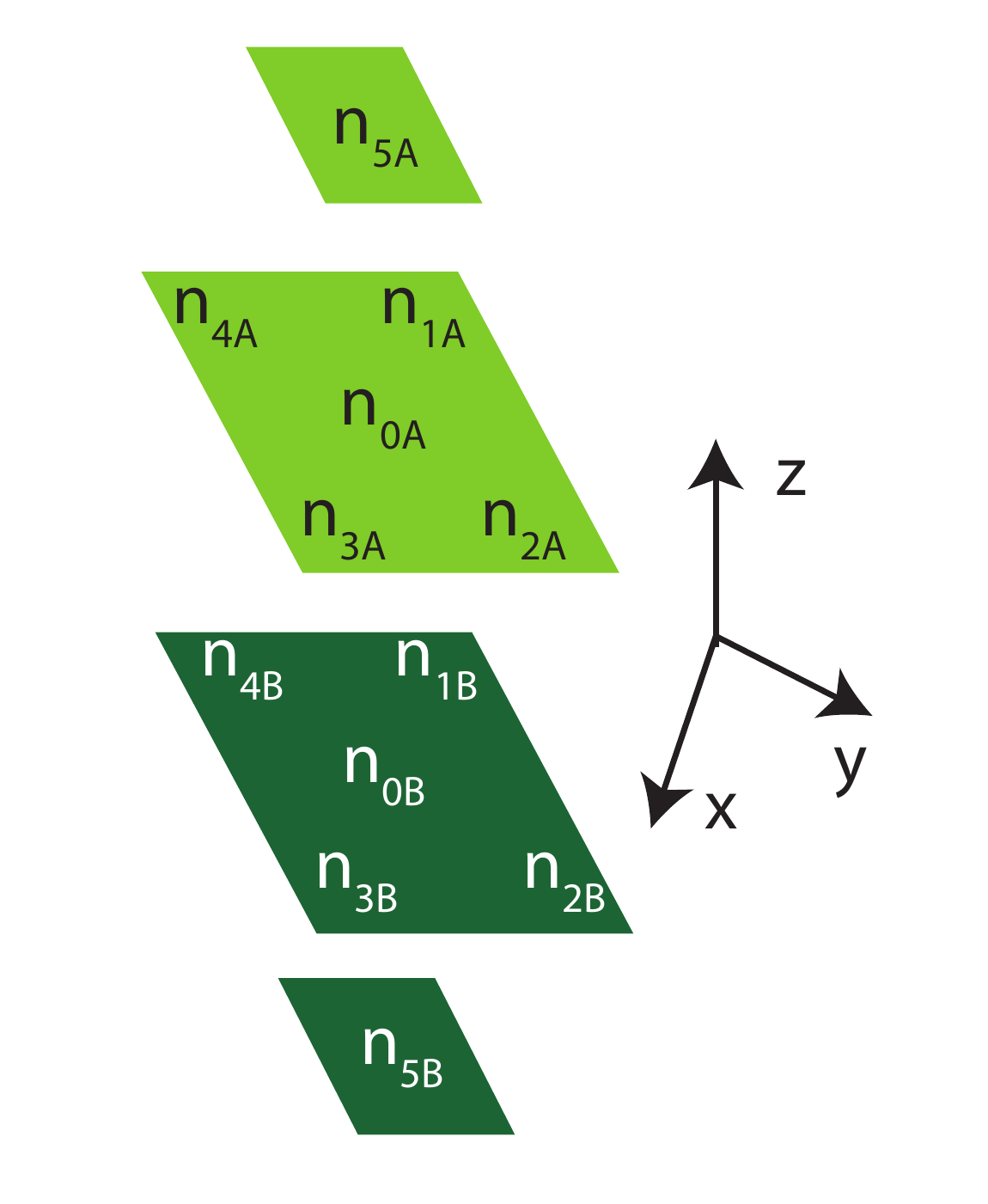}
\caption{\label{fig:cluster} Bethe-Peierls cluster schematic. $n_{0X}$ is the central site, and all others are neighboring sites. Dark-colored sites denote sites in a densely populated stripe, and light-colored sites represent sites in a sparsely populated stripe.}
\end{figure}

Our site cluster, depicted in Fig.~\ref{fig:cluster}, encompasses two thylakoid discs, so as to capture one instance of the striped motif in the striped phase. 

The cluster Hamiltonian is 

		\begin{align}
			\centering
			H &= - \mu (n_{0A} + n_{0B}) - J \sum_{i=1}^{4} (n_{0A} n_{iA} + n_{0B} n_{iB}) \nonumber \\
			&- \epsilon (n_{0A} n_{5A} + n_{0B} n_{5B}) + \epsilon' n_{0A} n_{0B} \nonumber \\ 
			&- \mu_A \sum_{i=1}^{4} n_{iA} - \mu_B \sum_{i=1}^{4} n_{iB} - \mu_{A}' n_{5A} - \mu_{B}' n_{5B}
		\end{align}
		where $A$ and $B$ denote different stripes. 	
	
	        In a BP ansatz, instead of solving for average densities, one solves for effective fields; these are given by $\mu_A$, $\mu_B$, $\mu_A'$, and $\mu_B'$. There are \textit{four} fields because sites interfacing with a stripe of the opposite type experience a different field than those surrounded by like sites. 
	
	 If we take $J = \epsilon$, then $\mu_{k}' = \mu_{k}$. With this in mind, we write the partition function. First, below are some important variable assignments:
		 
		\begin{align*}
			\mu_A &= \bar{\mu} + \Delta \mu,\mu_B = \bar{\mu} - \Delta \mu \nonumber \\
			z &= e^{\beta \mu}, z_A = e^{\beta \mu_A},z_B = e^{\beta \mu_B}\nonumber \\
			\bar{z} &= e^{\beta \bar{\mu}} ,\delta = e^{-\beta \epsilon'},c = e^{K} \nonumber \\
		\end{align*}
		
		Taking the standard derivatives of Eq.~\eqref{eq:bpq}, the average densities are Eq.~\eqref{eq:noa} and its $n_{0B}$ counterpart. Note that there are two average densities for each stripe, with $n_{0x}$ as the central sites and the others its neighboring sites.
		
	
	\begin{align} \label{eq:bpq}
			\centering
			Q &= \sum_{n_{0A},n_{0B}} z^{n_{0A} + n_{0B}} \delta^{n_{0A} n_{0B}} \bigg( 1 + z_A e^{K n_{0A}} \bigg)^5 \bigg( 1 + z_B e^{K n_{0B}} \bigg)^5 \nonumber \\
			&= \bigg( 1 + \bar{z} e^{\beta \Delta \mu} \bigg)^5 \bigg( 1 + \bar{z} e^{-\beta \Delta \mu} \bigg)^5 \nonumber \\ 
			&+ z \bigg( 1 + c \bar{z} e^{\beta \Delta \mu} \bigg)^5 \bigg( 1 + \bar{z} e^{-\beta \Delta \mu} \bigg)^5 \nonumber \\ 
			&+ z \bigg( 1 + \bar{z} e^{\beta \Delta \mu} \bigg)^5 \bigg( 1 + c \bar{z} e^{-\beta \Delta \mu} \bigg)^5 \nonumber \\ 
			&+ z^2 \delta \bigg( 1 + c \bar{z} e^{\beta \Delta \mu} \bigg)^5 \bigg( 1 + c \bar{z} e^{-\beta \Delta \mu} \bigg)^5 
	\end{align}

	\begin{align} \label{eq:noa}
			\langle n_{0A} \rangle &= \frac{1}{Q} \bigg( 1 + c \bar{z} e^{\beta \Delta \mu} \bigg)^5 z \bigg[ \bigg( 1 + \bar{z} e^{-\beta \Delta \mu} \bigg)^5 \nonumber \\
			&+ z \delta \bigg( 1 + c \bar{z} e^{-\beta \Delta \mu} \bigg)^5 \bigg]
	\end{align}
			
	\begin{align} \label{eq:na}
			\langle n_{A} \rangle &= \frac{1}{5Q} \frac{ \partial Q} { \partial \beta \mu_A} = \frac{z_A}{5Q} \frac{ \partial Q} { \partial z_A} \nonumber \\ 
			&= \frac{\bar{z} e^{\beta \Delta \mu}}{Q} \bigg\{ \bigg( 1 + \bar{z} e^{\beta \Delta \mu} \bigg)^4 \bigg[\bigg( 1 + \bar{z} e^{-\beta \Delta \mu} \bigg)^5 + z \bigg( 1 \nonumber \\ 
			&+ c \bar{z} e^{-\beta \Delta \mu} \bigg)^5 \bigg] 
			+ c z \bigg( 1 + c \bar{z} e^{\beta \Delta \mu} \bigg)^4 \bigg[ \bigg( 1 + \bar{z} e^{-\beta \Delta \mu} \bigg)^5 \nonumber \\ &+ z \delta \bigg( 1 + c \bar{z} e^{-\beta \Delta \mu} \bigg)^5 \bigg]	\bigg\}
	\end{align}
	
Here we have replaced $\mu_A$ and $\mu_B$ with $\bar{\mu} + \Delta \mu$ and $\bar{\mu} - \Delta \mu$, as this formulation more intuitively allows one to discuss the fields in terms of an average field and fluctuations from it. The astute reader will notice the factor of 5 in Eq.~\eqref{eq:na} -- this is the number of nearest neighbors in the same lattice $A$. In general, this number would be $2d - 1$, where $d$ is the total dimensionality of the system; other factors in Eq.~\eqref{eq:na} may change with different $d$. The difference between $\langle n_{iA} \rangle$ and $\langle n_{iB} \rangle$ simply involves replacing $\mu_A$ with $\mu_B$ and vice versa; for this reason, $\langle n_{iB} \rangle$ expressions are not shown here. 

Since we have two unknowns, $\bar{\mu}$ and $\Delta \mu$, instead of solving one self-consistency expression as for mean field theory, one must solve a system of equations. The system is Eqs.~\eqref{eq:syseqn1} or ~\eqref{eq:syseqn2}. The system was initialized for both small and large $\delta \mu$ and for initial $\mu_i$ large and small. The tolerance for self-consistency was $10^{-12}$, and the maximum number of iterations was 1 million. The transition was determined by finding $\Delta n$ differences larger than $10^{-9}$ between consecutive $\beta \mu$ values for a given $K$. 

	\begin{eqnarray} \label{eq:syseqn1}
		\langle n_{0A} \rangle &= \langle n_{A} \rangle \nonumber \\ 
		\langle n_{0B} \rangle &= \langle n_{B} \rangle 
	\end{eqnarray}
	
	\begin{eqnarray} \label{eq:syseqn2}
		\langle \Delta n_{0} \rangle - \langle \Delta n \rangle &= 0 \nonumber \\ 
		\langle \bar{n_{0}} \rangle + \langle \bar{n} \rangle &= 0 
	\end{eqnarray}
	where $\Delta n_{i} = \frac{1}{2} (n_A - n_B)$ and $\bar{ n_{i} }= \frac{1}{2} (n_A + n_B)$.

\subsection{Phase diagrams}

\begin{figure}
\centering
\includegraphics[width=0.75\textwidth]{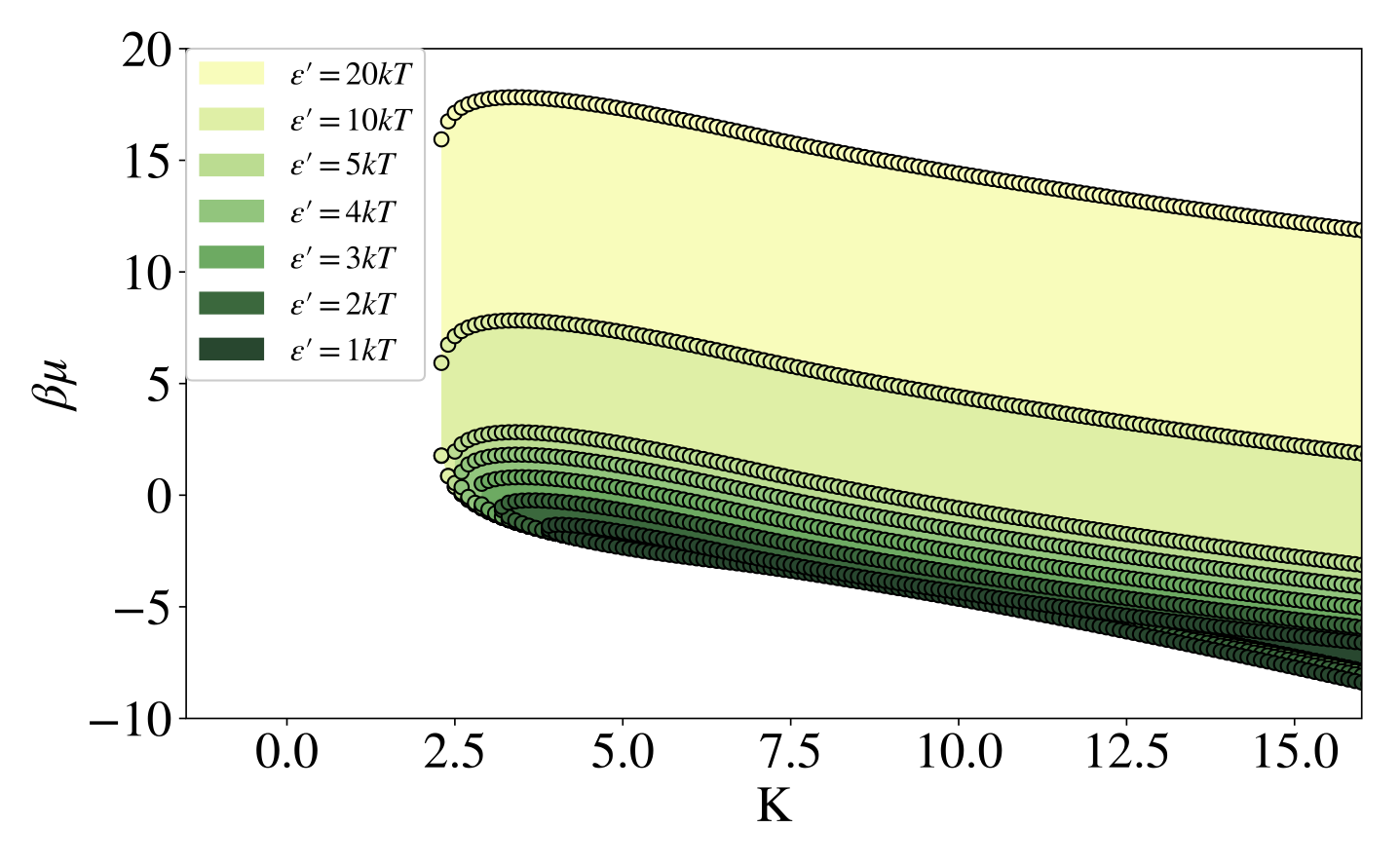}
\caption{\label{fig:bp1full} BP $\beta \mu$ vs. $K$ phase diagram, with first-order transitions beginning $K \approx 6 $. Shaded region indicates the striped phase. The upper branch was calculated via the inversion symmetry relation Eq. (4) in the main text.} 
\end{figure}

\begin{figure}
\centering
\includegraphics[width=0.75\textwidth]{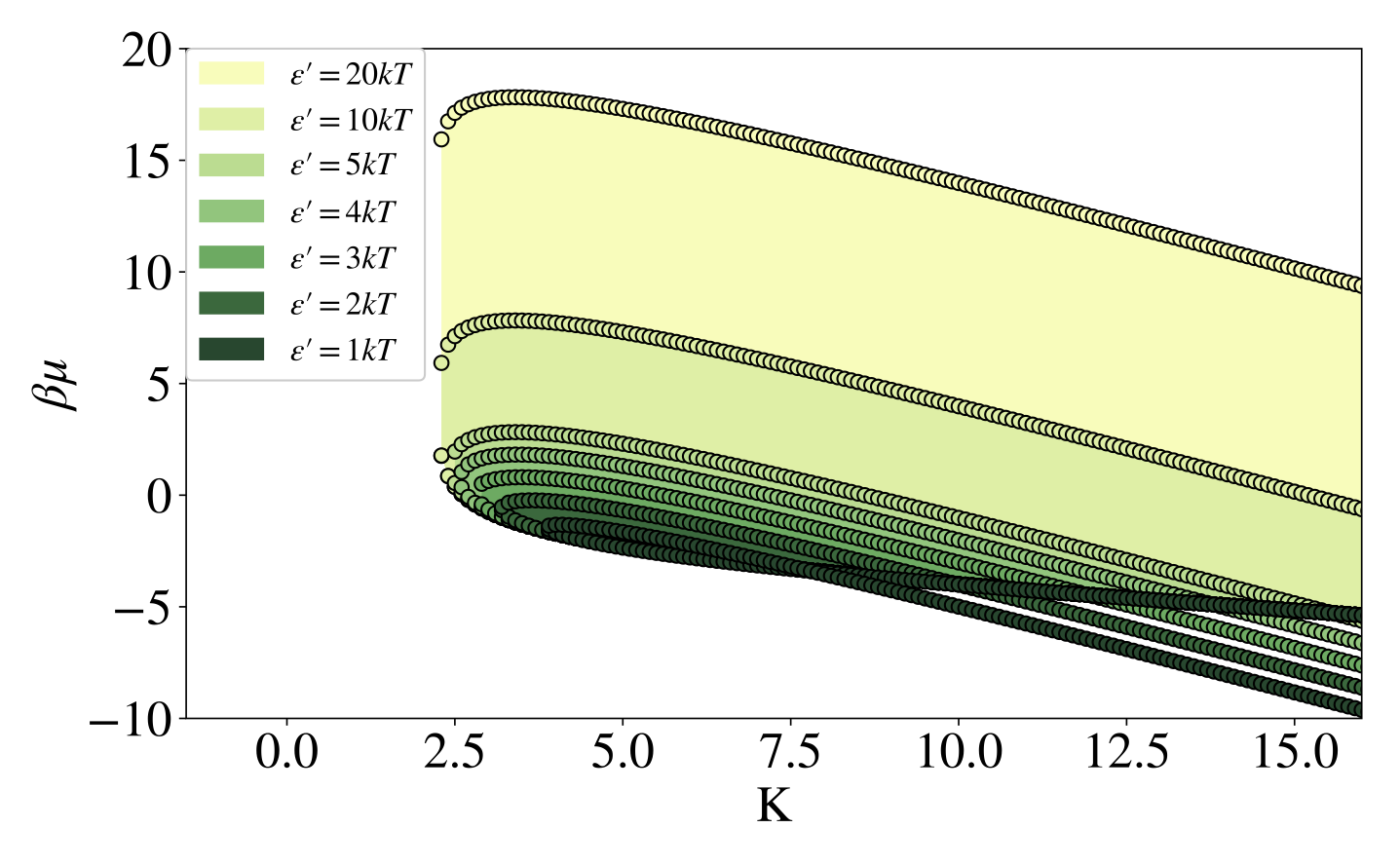}
\caption{\label{fig:bp2full} BP $\beta \mu$ vs. $K$ phase diagram, continuous transitions throughout. Shaded region indicates the striped phase. The upper branch was calculated via the inversion symmetry relation Eq. (4) in the main text.} 
\end{figure}

Here we present
Bethe-Peierls phase diagrams, in both the $\beta \mu$ vs. $K$ and
$\bar{n}$ vs. $K$ planes. Figs.~\ref{fig:bp1full} and \ref{fig:bp2full}
show a larger range of $K$ values than we presented
for two-site mean field theories. Only at these larger values of $K$
are signs of discontinuous ordering apparent at the BP level of mean
field theory.

\begin{figure}
\centering
\includegraphics[width=0.75\textwidth]{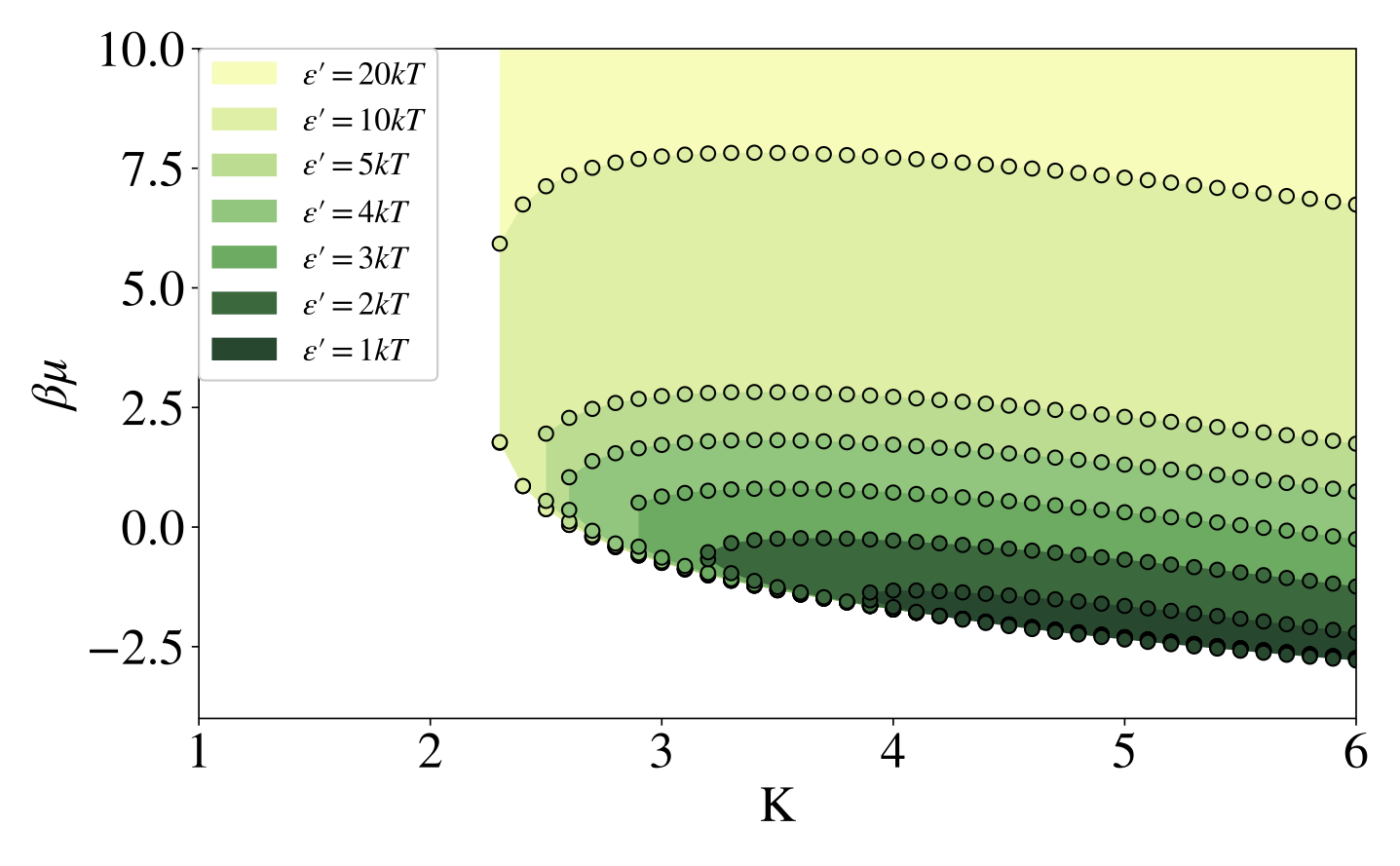}
\caption{\label{fig:bp2fulls} BP $\beta \mu$ vs. $K$ phase diagram, continuous transitions throughout. Shaded region indicates the striped phase. The upper branch was calculated via Eq. (4) in the main text.} 
\end{figure}

Continuous BP transitions can be determined by linearizing the
self-consistent equations. The resulting equations, which are
polynomial in $\bar{z}$, are amenable to numerical root finding
methods. Continuous transitions can also be located by initializing
the nonlinear self-consistent iteration appropriately. These
continuous transitions, plotted in Figs.~\ref{fig:bp2full} and
\ref{fig:bp2fulls} (on different scales), show the same unphysical crossing behavior found
with the two-site approach, though this crossing occurs at a larger
$K$ value than in the previous approach.

\begin{figure}
\centering
\includegraphics[width=0.75\textwidth]{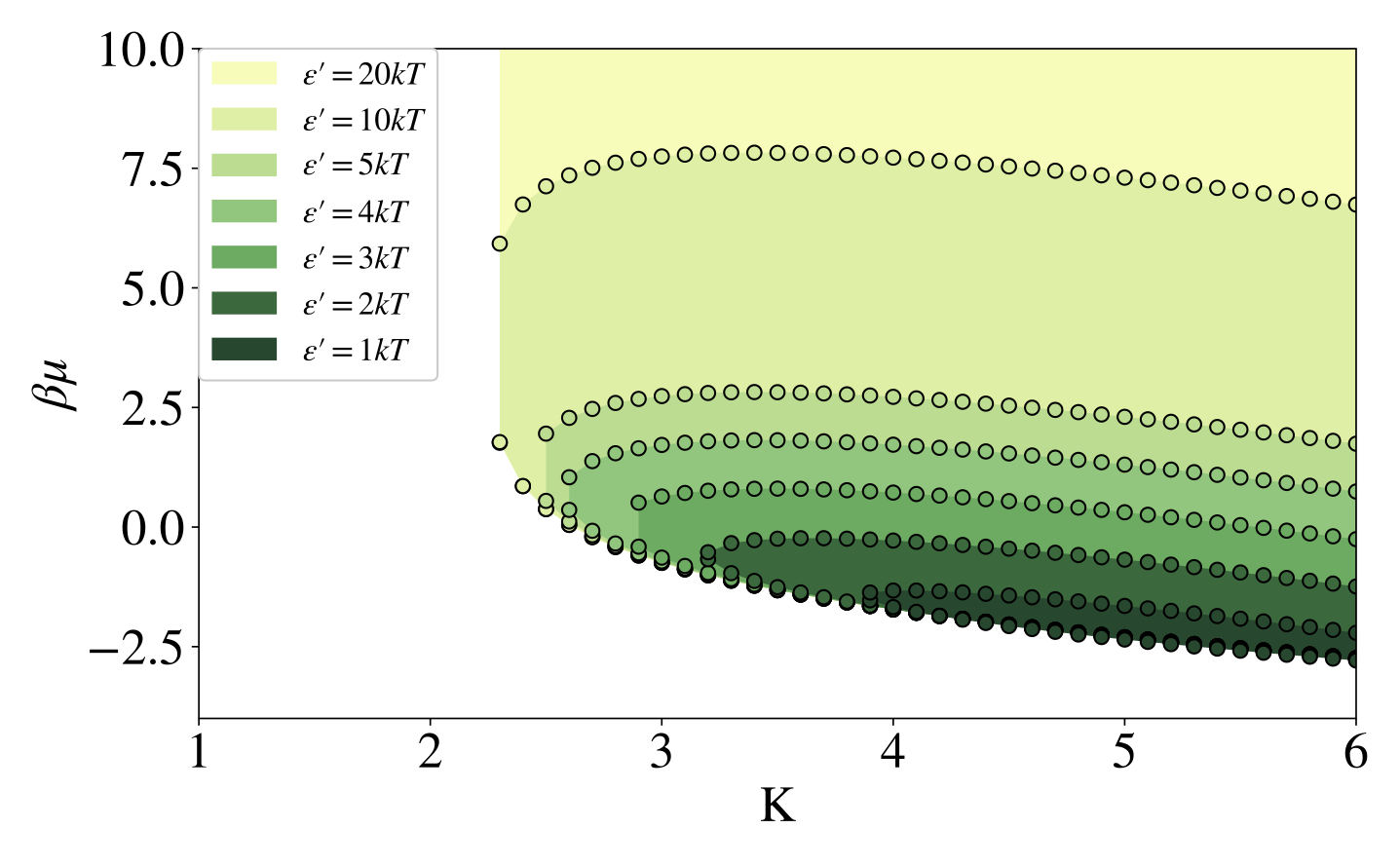}
\caption{\label{fig:bp1fulls} BP $\beta \mu$ vs. $K$ phase diagram, with first-order transitions beginning $K \approx 6 $ (not visible here). Shaded region indicates the striped phase. The upper branch was calculated via Eq. (4) in the main text.} 
\end{figure}

Self-consistent solutions obtained with a different
initialization are plotted in Fig.~\ref{fig:bp1full} over a wide range
of $K$. At small $K$ they coincide with the continuous transitions
described above, as emphasized in Fig.~\ref{fig:bp1fulls}, which shows
only the range of $K$ accessible in simulations. Limited to the domain
$1<K<6$, this plot is essentially identical to the continuous case
Fig.~\ref{fig:bp2fulls}. For large $K$, however, this initialization
produces different solutions, which do not cross. Instead, these phase
boundaries exhibit discontinuous change in both $\Delta n$ and
$\bar{n}$, and widen markedly at high $K$. All of these features are
consistent with results of two-site MF theory, but they set in at
higher $K$. For the range of $\epsilon'$ we have studied, the onset of
first-order transitions occurs near $K=6$, as opposed to the two-site
result of $K\approx 10/3$.
As per the data of Sec.~\ref{simcfirst}, 
first-order transitions are observed in simulation between $K=6$ 
and $K=7$, demonstrating that Bethe-Peierls does indeed more 
accurately estimate the location of discontinuous transitions in this model. 

The minimum value of $K$ at which ordering occurs is also shifted
upwards in BP theory, to about $K=2.4$. This prediction compares more
favorably with the critical value $K\approx 2.7$ found in simulations than
does the two-site prediction $K\approx 2$.

Again, viewed on 
the same scale as results in the main text, the BP
data very strongly resemble
the results of two-site mean field theory; see Figs.~\ref{fig:bp1fulls} to
\ref{fig:bp2fulls}. Note that these two figures are essentially
identical as the discontinuous transitions begin at $K \approx 6$.

\subsection{Two-cluster Bethe-Peierls approximation}

\begin{figure}[h]
\centering
\includegraphics[width=0.5\textwidth]{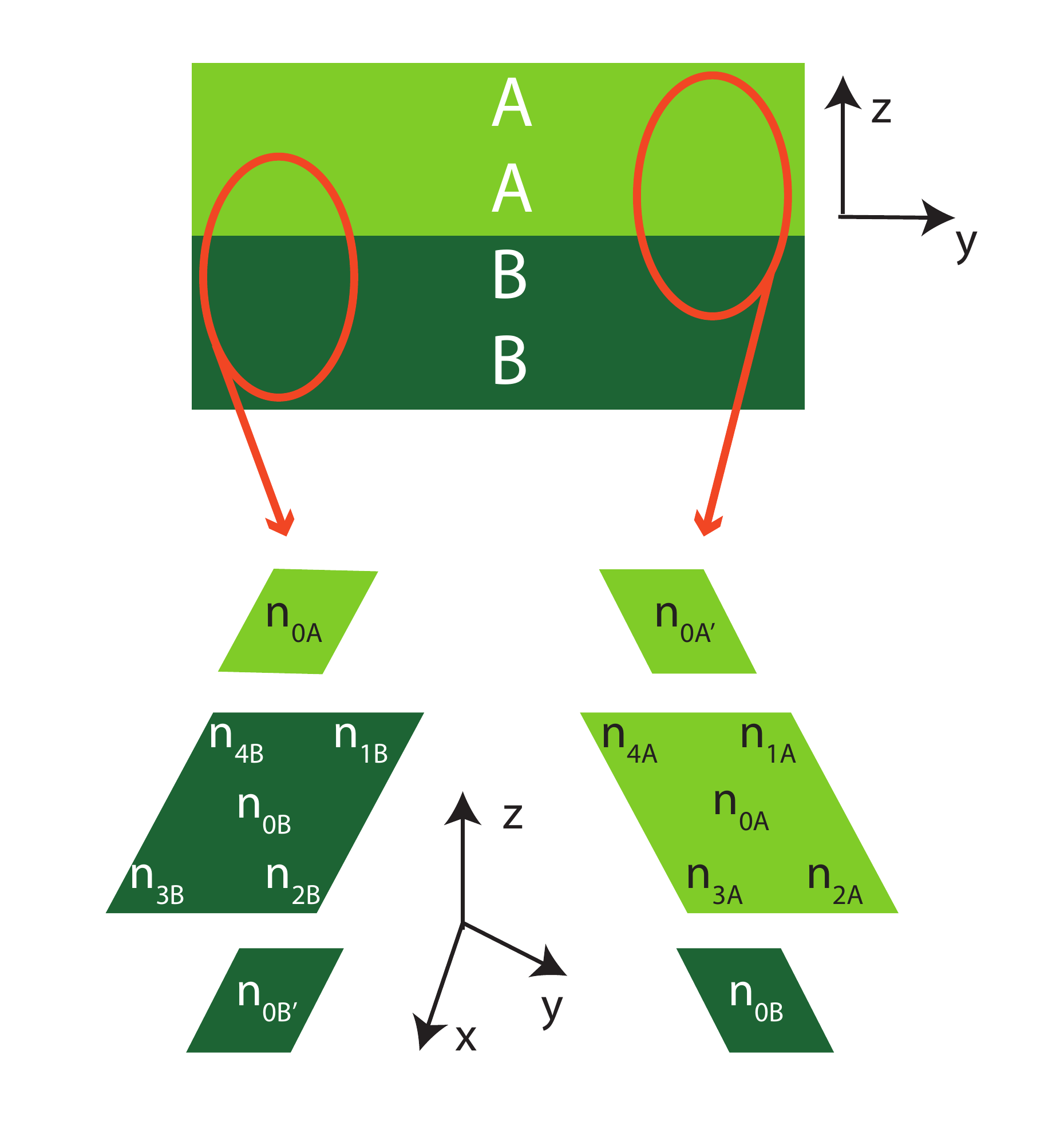}
\caption{\label{fig:acluster} Two-cluster schematic. The top bar represents four layers, with two sets of two-layer stripes. The lower half of the diagram represents the clusters used in the BP approximation, with $n_{0x}$ as the centers of the clusters.}
\end{figure} 

As mentioned in the main text, another way to account for alternating couplings in a lattice model using BP is to use two clusters instead of one (see Fig.~\ref{fig:acluster}). One cluster corresponds to a sparsely populated stripe, and the other a densely populated stripe. To start, we write the cluster Hamiltonian for cluster $A$: 

	\begin{equation}
		\begin{split}
			H_A &= -\mu n_{0A} -\mu_{B} n_{0B} - J n_{0A} (n_{1A} + n_{2A} + n_{3A} + n_{4A} ) \\
			&- \mu_{A} (n_{1A} + n_{2A} + n_{3A} + n_{4A} + n_{0A'}) \\
			& + \epsilon^{'} n_{0A} n_{0B} - \epsilon n_{0A} n_{0A'} 
		\end{split}
	\end{equation}
		
The cluster Hamiltonian for the $B$ lattice can be obtained similarly. The average densities arise in the traditional way, via derivatives of the partition function:
\\
\\
\\
\\
\begin{eqnarray} \label{eq:anoa}
	\langle n_{0A} \rangle &&= \frac{ \partial \ln Z_A} {\partial \beta \mu} 
		= \frac{1}{Z_A} e^{\beta \mu} \bigg( 1 + e^{\beta (\bar{\mu} + \Delta \mu + J)} \bigg)^4 \nonumber\\
		&&\times \bigg( 1 + e^{\beta (\bar{\mu} - \Delta \mu - \epsilon^{'})} \bigg) \bigg( 1 + e^{\beta (\bar{\mu} + \Delta \mu + \epsilon)} \bigg) 
\end{eqnarray}

\begin{widetext}
\begin{eqnarray} \label{eq:ana}
	\langle n_{A} \rangle &&= \frac{1}{5} \frac{ \partial \ln Z_A} {\partial \beta \mu_A} \nonumber \\
	&&= \frac{1}{5 Z_A} \bigg[
	5 \bigg( 1 + e^{\beta (\bar{\mu} + \Delta \mu}) \bigg)^4 e^{\beta (\bar{\mu} + \Delta \mu)} \bigg( 1 + e^{\beta (\bar{\mu} - \Delta \mu)} \bigg) \nonumber \\ 
	&&+ 4 \bigg( 1 + e^{\beta (\bar{\mu} + \Delta \mu + J)} \bigg)^3 e^{\beta (\bar{\mu} + \Delta \mu + J)} e^{\beta \mu} \bigg( 1 + e^{\beta (\bar{\mu} - \Delta \mu - \epsilon^{'}) } \bigg) 
	 \bigg( 1 + e^{\beta (\bar{\mu} + \Delta \mu + \epsilon) } \bigg) \nonumber \\
	 &&+ e^{\beta \mu} e^{\beta (\bar{\mu} + \Delta \mu + \epsilon) } \bigg( 1 + e^{\beta (\bar{\mu} - \Delta \mu - \epsilon^{'}) } \bigg) 
	\bigg( 1 + e^{\beta (\bar{\mu} + \Delta \mu + J)} \bigg)^4 \bigg]
\end{eqnarray}
\end{widetext}

Using these expressions, the same systems of equations 
~\eqref{eq:syseqn1} or \eqref{eq:syseqn2} were solved via gradient
descent optimization to find continuous transitions. Please note that
no first-order transitions were found for this model, and nor did
these equations preserve occupation inversion symmetry.

\subsubsection{Momentum-boosted gradient descent}
The system of equations, Eq.~\eqref{eq:syseqn1}, was rephrased as a root-finding problem in Eq.~\eqref{eq:syseqn2}, such that a gradient descent method could be used to find its roots. Consequently, one can imagine the system of equations as a vector whose components are the equations. Thus, the objective function optimized was the magnitude of this vector -- namely, the sum of the squared equations set equal to zero. 

A gradient descent approach was used for a number of reasons. Firstly, the ideal initial conditions for this system were unknown, so a method that can handle initial conditions far from the solution was desired; many root-finding and optimization algorithms do not do well when seeded far from the solution. Second, so as to handle potentially multiple solutions for a given set of parameters, we wanted a method that had the ability to find multiple minima -- this concern is related to the first, since initial conditions must be given differently so as to explore possible multiple global solutions. Gradient descent is algorithmically simple and has mostly guaranteed convergence, hence it was chosen. 

Furthermore, MGD was utilized instead of plain steepest descent as a means of increasing efficiency and preventing traps in local optima. \cite{Ruder2017ar} One can write the $x$ component of the update vector at the next step as 

\begin{eqnarray} \label{mgd}
v_{x,t+1} &= \gamma v_{x,t} + s \nabla_x f \nonumber \\ 
x &= x - v_{x,t+1} 
\end{eqnarray}

where $\gamma$ is the momentum scalar that usually is between 0.9 and 1 and encodes the "memory" of the previous step, and $s$ is the step size for the descent. The step size is on the order of 0.1 to 0.0001 usually. Convergence was determined by how close both the objective function and the gradient were to zero. The gradient of Eq.~\eqref{eq:syseqn2} was approximated using finite differences. 

For each system of BP equations, the initial conditions were generated by creating a grid of ($\Delta \mu$, $\bar{\mu}$) values. Since $\bar{\mu}$ was expected to remain relatively close to $\mu$, a limited number of $\bar{\mu}$ initial guesses were used. For $\Delta \mu$, a grid ranging from -2 to 2 $k_B T$ measured out by a given increment were used; based on preliminary explorations, solutions obeying the constraints of the problem are only found for relatively small $\Delta \mu$ (that is, $\Delta \mu$ within these bounds). 



\end{document}